\documentclass[12pt]{article}

\usepackage[T1]{fontenc}
\usepackage[utf8]{inputenc}
\usepackage{lmodern} 
\usepackage{amsmath,amssymb}
\usepackage{graphicx}
\usepackage{booktabs}
\usepackage{siunitx}
\usepackage{authblk} 
\usepackage[top=0.75in,bottom=0.75in,left=0.75in,right=0.75in]{geometry}
\usepackage[font=footnotesize]{caption}
\usepackage{subcaption} 
\usepackage{hyperref} 
\usepackage{soul}

\title{Nonreciprocal Negative Refraction Enabled by Photonic Time Crystals}

\author[1]{Mohammad R. Tavakol}
\author[1,2,*]{Wenshan Cai}
\affil[1]{School of Electrical and Computer Engineering, Georgia Institute of Technology, Atlanta, Georgia 30332, United States}
\affil[2]{School of Materials Science and Engineering, Georgia Institute of Technology, Atlanta, Georgia 30332, United States}
\affil[*]{Correspondence: \texttt{wcai@gatech.edu}}

\date{} 

\begin{document}
\maketitle

\begin{abstract}
We propose and theoretically demonstrate nonreciprocal negative refraction enabled by time-varying photonic structures. By engineering temporal modulations at the interfaces of hyperbolic media, we achieve isolation between forward and backward beams while preserving the hallmark property of negative refraction. Two complementary approaches are developed: in the optical regime, a multilayer AZO/ZnO hyperbolic slab is sandwiched between permittivity-modulated dielectric layers (3D time crystals); in the microwave regime, a wire medium is sandwiched between time-modulated resistive metasurfaces (2D time crystals). Both designs exploit Floquet harmonic expansions and are validated with a customized harmonic-balance finite-element solver. We report isolation exceeding 46 dB in the optical device and ~11 dB in the microwave counterpart. This work introduces a general framework for nonreciprocal negative refraction across frequency regimes, expanding the design space of time-varying metasurfaces and photonic time crystals.
\end{abstract}



\section{Introduction}

The concept of negative refraction has long fascinated the photonics community because of its ability to fundamentally alter the way electromagnetic waves propagate. Following Veselago’s theoretical predictions, experimental realizations of negative-index metamaterials enabled unprecedented control over light propagation and established new regimes of optics \cite{poddubny_hyperbolic_2013,huo_hyperbolic_2019,shalaev_negative_2005}. Negative refraction enables a range of unusual wave phenomena, including reversed Cherenkov radiation, negative Goos–Hänchen shifts, backward-wave propagation, and negative radiation pressure \cite{cai_optical_2010}. Among the various approaches to realizing negative refraction, \textit{hyperbolic media} stand out because of their highly anisotropic dispersion relations, which support both propagating and evanescent modes and give rise to backward-wave refraction \cite{yao_optical_2008}. These properties have been exploited in multilayer plasmonic and semiconductor systems, including transparent conducting oxide multilayers such as aluminum-doped zinc oxide (AZO)/ZnO stacks, which offer tunability and compatibility with nanophotonic platforms \cite{naik_demonstration_2012}.  

Despite such progress, conventional negative refraction remains fundamentally \textit{reciprocal}. Lorentz reciprocity dictates that under linear and time-invariant conditions, electromagnetic systems respond identically to forward and backward excitation\cite{shaltout_time-varying_2015}. Thus, a beam incident from the forward direction experiences the same refraction angle and transmission efficiency as one entering from the backward direction, which prevents any inherent directionality in the response.

The traditional route to breaking reciprocity is through magneto-optical effects, in which time-reversal symmetry is broken by applying an external magnetic bias \cite{fan_magneto-optical_2019}. Magneto-optical devices are widely used in microwave engineering and optical communication, yet they suffer from severe drawbacks: they are bulky, require strong magnetic fields, and are challenging to integrate on-chip. This has fueled intense efforts to achieve magnetic-free nonreciprocity \cite{fan_nonreciprocal_2018}.  

Among various approaches, \textit{temporal modulation} has emerged as a particularly promising strategy \cite{yu_complete_2009,nagulu_non-reciprocal_2020,shi_optical_2017,dinc_synchronized_2017,khurgin_optical_2023}. Temporally varying permittivity, permeability, or conductivity breaks time-reversal symmetry directly and couples the incident frequency to sidebands at harmonics of the modulation frequency \cite{sounas_non-reciprocal_2017}. Unlike static anisotropy or geometric asymmetry, which preserve reciprocity, temporal modulation induces frequency conversion and directional asymmetry, since forward and backward waves interact with the modulation phase differently. Related transmissive time-modulated media and space–time Bragg gratings have been shown to exhibit robust optical nonreciprocity and compact nonreciprocal responses.\cite{ramaccia_electromagnetic_2020,sedeh_optical_2022,guo_nonreciprocal_2019} Theoretically, even uniform time modulation in bianisotropic systems can produce nonreciprocity under appropriate coupling conditions, reinforcing the generality of temporal-symmetry breaking \cite{wang_nonreciprocity_2020}.

The concept of \textit{photonic time crystals (PTCs)} --- systems with periodic modulation in time --- naturally extends photonic crystals into the temporal domain. Just as spatial periodicity produces photonic bandgaps, temporal periodicity creates frequency bandgaps and exotic dispersion properties \cite{lustig_topological_2018}. Theoretical work has predicted amplification, frequency conversion, and temporal Bragg scattering in such systems \cite{galiffi_photonics_2022, engheta_four-dimensional_2023}. Recent experiments have confirmed aspects of these predictions, showing the feasibility of dynamic index modulation on sub-cycle timescales \cite{asgari_theory_2024}. Diverse mechanisms have been explored: carrier depletion in semiconductors\cite{lira_electrically_2012}, photon acceleration in transparent conducting oxides \cite{jaffray_transparent_2022}, and time-varying surface impedances in microwaves\cite{wang_metasurface-based_2023}. For example, varactor-loaded or resistive patch arrays have been used to implement time-varying admittances, producing strong nonreciprocal responses in compact microwave devices \cite{wu_serrodyne_2020,mostafa_coherently_2022}. Theoretical frameworks based on generalized transfer matrices \cite{ramaccia_temporal_2021}, modal methods\cite{galiffi_wood_2020,taravati_generalized_2019}, and transmission-line circuit models \cite{wang_metasurface-based_2023, wang_theory_2020,ramaccia_phase-induced_2020} have unified the treatment of such temporal and spatiotemporal systems, further broadening their applicability. Additionally, parallel advances in spatiotemporal metasurfaces further highlight the utility of temporal modulation\cite{hadad_space-time_2015,zhang_space-time-coding_2018, harwood_space-time_2025}. These platforms combine spatial structuring with time modulation to realize asymmetric transmission, and frequency conversion\cite{wang_theory_2020,zang_nonreciprocal_2019, galiffi_wood_2020, zhang_breaking_2019,tirole_second_2024}.   

Despite these advances, \textit{nonreciprocal negative refraction} has not been achieved. Existing works fall into two distinct categories: negative refraction in static hyperbolic or negative-index systems (reciprocal) \cite{poddubny_hyperbolic_2013, naik_demonstration_2012}, and nonreciprocity in temporally modulated systems (without negative refraction) \cite{sounas_non-reciprocal_2017, mostafa_coherently_2022, chegnizadeh_non-reciprocity_2020}. Achieving directional control in systems that refract light negatively would expand the functionality of negative-index and hyperbolic platforms, enabling asymmetric beam steering, isolating components, and robust signal-routing capabilities that static negative-refraction systems fundamentally lack. Combining the two requires careful integration of hyperbolic dispersion with temporal modulation to maintain negative refraction while breaking reciprocity.

In this work, we present the demonstration of nonreciprocal negative refraction enabled by time crystals. Our approach leverages temporal modulation as a symmetry-breaking tool that enables nonreciprocal control of negatively refracted beams. To implement this concept, we sandwich a hyperbolic medium between two temporally modulated interfaces driven with a quadrature-phase offset (90$^\circ$). This configuration ensures that forward and backward beams encounter distinct temporal states of the system. As a result, both directions undergo negative refraction, but their transmission amplitudes differ strongly, yielding isolation. Together, these features unite the physics of hyperbolic dispersion with symmetry breaking from time modulation, establishing a general framework for nonreciprocal negative refraction.

We exemplify the concept in two complementary frequency regimes. In the optical regime, we design a multilayer AZO/ZnO hyperbolic stack bounded by permittivity-modulated dielectric slabs, which act as \textit{3D time crystals}. This system leverages reactive modulation, where energy is stored and redistributed among harmonics. Rigorous theory and simulations confirm isolation exceeding 46 dB while maintaining negative refraction. In the microwave regime, we use a wire medium bounded by resistive metasurfaces whose sheet conductances are temporally modulated with a quadrature-phase offset. This represents a \textit{2D time crystal} that leverages absorptive modulation, where direction-dependent dissipation enforces nonreciprocity. The achieved isolation is $\sim$11 dB, smaller than in the optical case, but the microwave implementation permits much stronger temporal modulation, which is easier to realize in conductance-modulated metasurfaces than in optically modulated dielectrics.

Together, these two platforms exhibit the universality of the framework. By integrating hyperbolic dispersion with time-crystal interfaces, we achieve a previously unrealized form of nonreciprocal negative refraction. This approach opens pathways toward magnet-free isolators, asymmetric beam routing devices, and integrated nonreciprocal components, while also extending the physics of photonic time crystals into new regimes.

\section{Results and Discussion}

\subsection{Conceptual Framework}

\begin{figure*}[!t]
\centering
\includegraphics[width=0.95\linewidth]{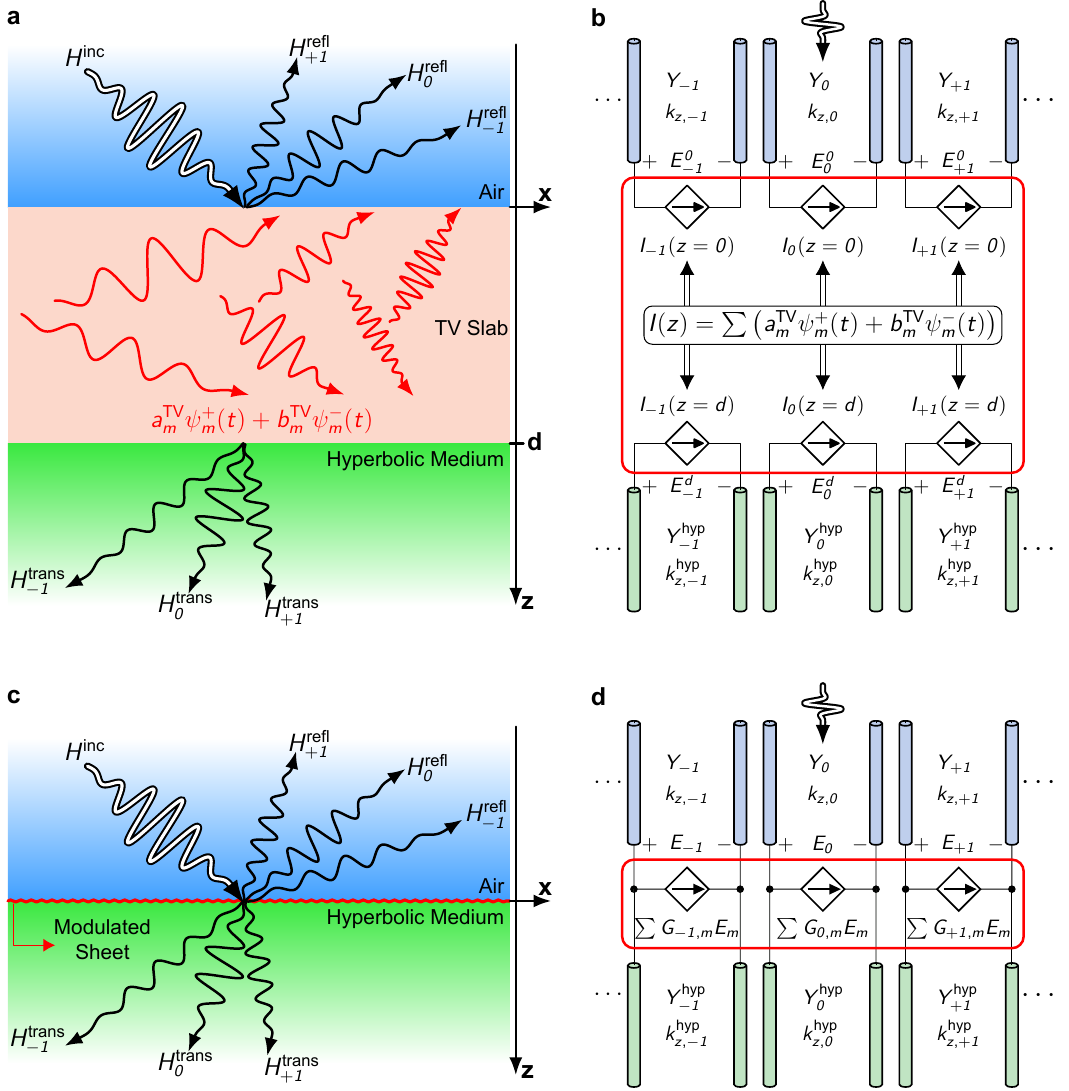}
\caption{Modeling time-varying interfaces through negative refraction. 
(a) Schematic of a multilayer structure where a TM-polarized wave impinges from air (blue) onto a time-varying (TV) dielectric slab of thickness $d$ (red) adjacent to a hyperbolic medium (green). The incident beam (thick white arrow) excites reflected and transmitted harmonics (black arrows). Inside the TV slab, the field is a superposition of forward and backward multi-harmonic modes, $\psi_m^{+}(t)$ and $\psi_m^{-}(t)$, with amplitudes $a_m^{\text{TV}}$ and $b_m^{\text{TV}}$. Different harmonics of these modes are shown by the red arrows. Black and red wavy arrows with varying oscillation periods denote distinct harmonics. The structure is homogeneous along the $y$-axis. 
(b) Equivalent circuit model of (a). Each harmonic in the air and hyperbolic regions is represented by a transmission line with propagation constant $k_{z,m}$ and wave admittance $Y_m$. The incident excitation couples into the zeroth-order line of the air region. The TV slab is modeled (red box) as a linear combination of forward and backward harmonics, with voltage-controlled current sources mediating inter-harmonic coupling. Three dots before and after the lines indicate that only a subset of the infinite harmonic spectrum is shown. 
(c) Schematic of a configuration with a time-modulated conductive sheet (red line) at the air--hyperbolic interface. Unlike the TV slab in (a), the sheet generates multi-harmonic surface currents that couple the incident excitation to reflected and transmitted harmonics. Similar to (a), the structure is homogeneous along the $y$-axis. 
(d) Equivalent circuit model of (c). Transmission lines again represent harmonics in the air and hyperbolic regions. The time-modulated sheet is modeled (red box) by shunt admittances realized through voltage-controlled current sources, which directly couple harmonics across the interface.}
\label{fig:fig1}
\end{figure*}

We begin by outlining the general concept of time-crystal-enabled nonreciprocal negative refraction. Figure~\ref{fig:fig1} illustrates the core idea. A TM-polarized plane wave at frequency $\omega_0$ impinges on a hyperbolic slab that is bounded on both sides by time-varying interfaces. Because of temporal modulation, the incident frequency couples to sidebands $\omega_m = \omega_0 + m\Omega$, exciting a spectrum of harmonics in both reflection and transmission. These two interfaces with distinct temporal phases will be employed to break the reciprocity of negative refraction through the hyperbolic region.

For a 3D time crystal region as the interface, the modulation of permittivity in the dielectric slab regions is described by $\epsilon_\ell(t) = \epsilon_{r,0} + \Delta \epsilon \cos(\Omega t + \phi_\ell), \ \ell = 1,2$, representing the top and the bottom slabs. As shown schematically in Fig.~\hyperref[fig:fig1]{\ref*{fig:fig1}(a)}, the incident TM wave excites forward- and backward-propagating multiharmonic eigenmodes inside the time-varying (TV) slab. These are denoted $\psi_m^+(t) = \psi_m^{\mathrm{TV}}(t)e^{-j k^{\mathrm{TV}}_{z,m} z}$ and $\psi_m^-(t) = \psi_m^{\mathrm{TV}}(t)e^{+j k^{\mathrm{TV}}_{z,m} z}$, with amplitudes $a_m^{\mathrm{TV}}$ and $b_m^{\mathrm{TV}}$, respectively. Their superposition can be written as $\sum_m \left( a_m^{\mathrm{TV}} \psi_m^+(t) + b_m^{\mathrm{TV}} \psi_m^-(t) \right)$. These modes mediate coupling into reflected and transmitted harmonics in the air and hyperbolic regions, indicated by the black wavy arrows of different periods. In this way, the TV slab generates multiple frequency channels, each corresponding to a Floquet harmonic.

The equivalent circuit model in Fig.~\hyperref[fig:fig1]{\ref*{fig:fig1}(b)} provides a compact representation. Each harmonic $m$ in the air region is modeled by a transmission line with propagation constant $k_{z,m} = \sqrt{(\omega_m/c)^2 - k_{x,\mathrm{inc}}^2}$, where $\omega_m = \omega_0 + m\Omega$, and characteristic admittance $Y_m = (k_m/k_{z,m})(1/\eta)$ with $k_m = \omega_m/c$ and $\eta$ the free-space impedance. Similarly, each harmonic in the hyperbolic region is represented by a line with $k^{\mathrm{hyp}}_{z,m} = \sqrt{\epsilon_t \left( \omega_m^2/c^2 - k_{x,\mathrm{inc}}^2/\epsilon_z \right)}$ and $Y_m^{\mathrm{hyp}} = (k_m/k^{\mathrm{hyp}}_{z,m})(1/\eta)$. The red box in Fig.~\hyperref[fig:fig1]{\ref*{fig:fig1}(b)} represents the TV slab, which is modeled as a set of voltage-controlled current sources at $z=0$ and $z=d$. Each source is a linear combination of the harmonic electric field amplitudes at the slab boundaries, $E^0_m$ and $E^d_m$, and is determined by enforcing field boundary conditions. In this picture, the transmission lines represent the time diffraction orders, while the controlled sources capture harmonic mixing induced by modulation.

For a 2D time crystal boundary as the interface, as shown in Fig.~\hyperref[fig:fig1]{\ref*{fig:fig1}(c)},the time-varying interfaces are resistive or conductive sheets whose conductance is modulated as $\sigma_\ell(t) = \sigma_0 + \Delta \sigma \cos(\Omega t + \phi_\ell), \ \ell=1,2$. Here, the mechanism is similar but simpler. The sheet generates surface currents that directly couple the incident excitation to reflected and transmitted harmonics. In the circuit representation (Fig.~\hyperref[fig:fig1]{\ref*{fig:fig1}(d)}), this appears as shunt admittances (red box) that connect harmonics in the air and hyperbolic regions. The coupling currents are determined by convolution of the Toeplitz matrix associated with $\sigma_\ell(t)$ with the boundary field amplitudes $E_m$. Thus, while the slab case involves distributed volumetric modulation, the sheet case involves surface currents that directly enforce asymmetric coupling across the interface.

The thick white arrow in Figs.~\hyperref[fig:fig1]{\ref*{fig:fig1}(a)} and \hyperref[fig:fig1]{\ref*{fig:fig1}(c)} denotes the oblique incident wave at frequency $\omega_0$, while the thinner wavy arrows denote reflected and transmitted harmonics. In both the slab and sheet configurations, the key principle is that time modulation produces a ladder of Floquet harmonics, some propagating and others evanescent. With a $\pi/2$ phase offset between the two modulated boundaries, these harmonics interfere constructively in one direction (forward) and destructively in the other (backward). This interference results in strong direction-dependent transmission while retaining the negative refraction associated with the hyperbolic core.

Taken together, Fig.~\ref{fig:fig1} and the above formulations show how time-varying boundaries on a hyperbolic medium transform reciprocal negative refraction into a nonreciprocal effect. In the optical design, the modulated slabs act as 3D photonic time crystals where volumetric modulation excites multiharmonic eigenmodes. In the microwave design, the modulated sheets act as 2D photonic time crystals where surface currents enforce asymmetric coupling. Both approaches implement the same principle: interference between Floquet harmonics controlled by quadrature modulation, yielding nonreciprocal negative refraction. The following subsections illustrate these principles concretely in the optical (Fig.~\ref{fig:fig2}) and microwave (Fig.~\ref{fig:fig4}) regimes.

\subsection{Optical Device: 3D Time Crystals}

We first consider the optical implementation, which employs a multilayer AZO/ZnO hyperbolic slab sandwiched between two permittivity-modulated dielectric layers. AZO is a transparent conducting oxide with metallic response in the near-IR, while ZnO serves as the dielectric partner. When alternated in thin layers, the composite exhibits hyperbolic dispersion, enabling negative refraction for TM waves.  

\begin{figure*}[!t]
\centering
\includegraphics[width=0.99\linewidth]{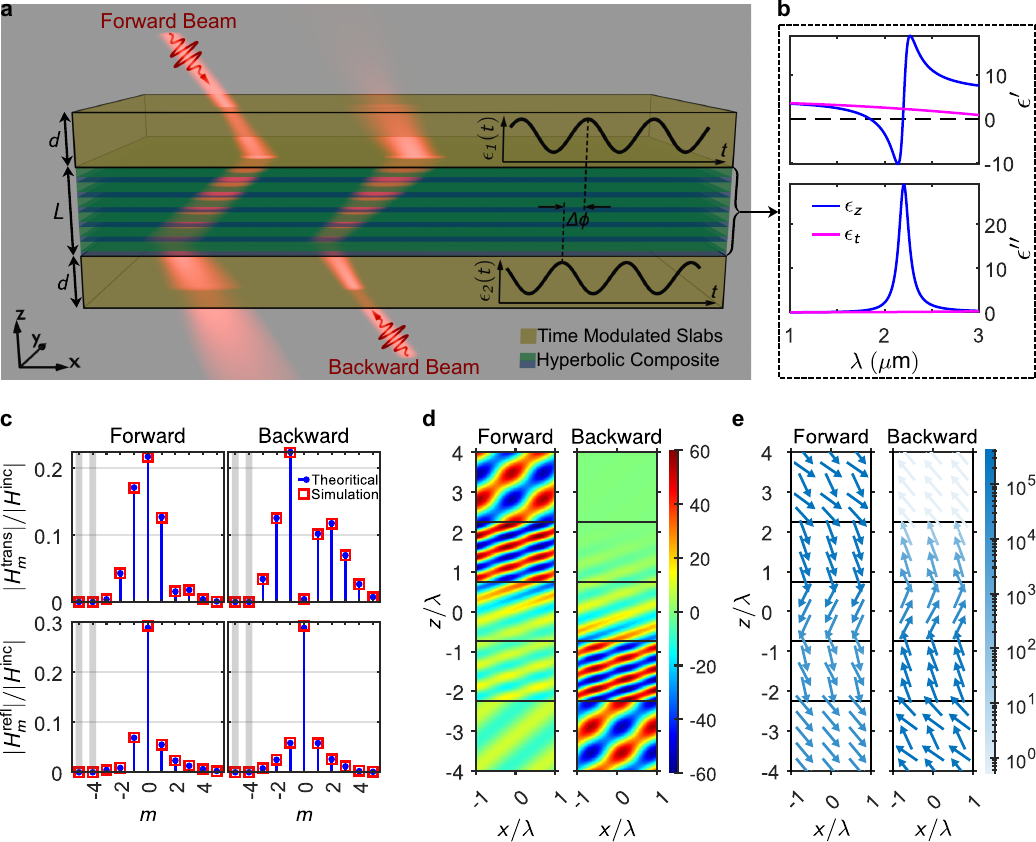}
\caption{Working principle and typical response of the proposed nonreciprocal negative refraction in the optical range. (a) A schematic of the structure composed of an alternating layer stack of thin plasmonic and dielectric films (AZO/ZnO), sandwiched between two dielectric slabs with modulated permittivity. Besides the subwavelength features of the hyperbolic layer, the system is mainly characterized by the hyperbolic slab thickness, $L = 2.8 \ {\rm{ \mu m}}$, and the incident angle, $\theta_\mathrm{inc} = 40^\circ$. The permittivities of the two modulated slabs are varied with the same frequency and modulation depth but have different phases, i.e., $\phi_1$ and $\phi_2$. The forward and backward oblique beams exhibit significantly different transmissions, resulting in nonreciprocal behavior. The red light beams schematically depict diffraction (due to harmonic modulations) and refraction through the structure. (b) The real and imaginary parts of the effective perpendicular and transverse permittivities of the thin film stack, used as the hyperbolic medium to realize negative refraction. The periodicity of the thin films along the structure is subwavelength, with a metal filling factor of $f = t_m / (t_m + t_d) = 0.3$, where $t_m$ and $t_d$ are the thicknesses of the metal and dielectric layers, respectively. (c) Forward and backward transmission and reflection field amplitudes, obtained through simulation and analytical calculation, versus harmonic order, $m$, with each frequency harmonic $\omega_m$ being $ \omega_0 + m\Omega$. (d) The magnetic field profile ($Hy$) when the structure is illuminated from the top (forward) and bottom (backward). Both surface plots use the same linear scale, shown via the colorbar on the right. (e) Power vector fields corresponding to the field profile in (d). The vector fields are normalized, and the arrow colors indicate power amplitude. Both vector field plots share the same linear scale, represented by the colorbar on the right.}
\label{fig:fig2}
\end{figure*}

The structure is shown schematically in Fig.~\hyperref[fig:fig2]{\ref*{fig:fig2}(a)}. The hyperbolic dispersion relation, calculated via effective medium theory (see Supplementary), is plotted in Fig.~\hyperref[fig:fig2]{\ref*{fig:fig2}(b)}, revealing Type-II hyperbolicity near $\lambda = 1.9~\mu$m. These multilayers have been widely studied as tunable hyperbolic metamaterials \cite{poddubny_hyperbolic_2013}, and their integration with time-modulated interfaces marks an unexplored frontier in coupling hyperbolic dispersion and temporal modulation across distinct regions.

Figures~\hyperref[fig:fig2]{\ref*{fig:fig2}(c)--(e)} present the central results. Under forward incidence, the zeroth-order transmission dominates, and the beam undergoes negative refraction with moderate insertion loss. Under backward incidence, by contrast, the zeroth-order transmission is strongly suppressed. Figure~\hyperref[fig:fig2]{\ref*{fig:fig2}(c)} plots the harmonic spectra, where forward illumination yields significant zeroth-order transmission, while backward illumination is nearly extinguished. Field maps of $H_y$ in Fig.~\hyperref[fig:fig2]{\ref*{fig:fig2}(d)} show a clear negatively refracted beam for forward incidence, while the backward case produces only weak scattered fields. Poynting vector distributions in Fig.~\hyperref[fig:fig2]{\ref*{fig:fig2}(e)} confirm the effect, revealing direction-dependent energy flow through the structure.  

We quantify the asymmetry using the isolation metric
\begin{equation}
I = 10 \log_{10}\!\left(\frac{T_f}{T_b}\right),
\label{eq:isolation}
\end{equation}
where $T_f$ and $T_b$ are the forward and backward zeroth-order power transmission coefficients. At the operating wavelength, we obtain $I \approx 46.4$ dB, establishing strong nonreciprocity while preserving negative refraction.  

\begin{figure}[!t]
\centering
\includegraphics[width=0.47\linewidth]{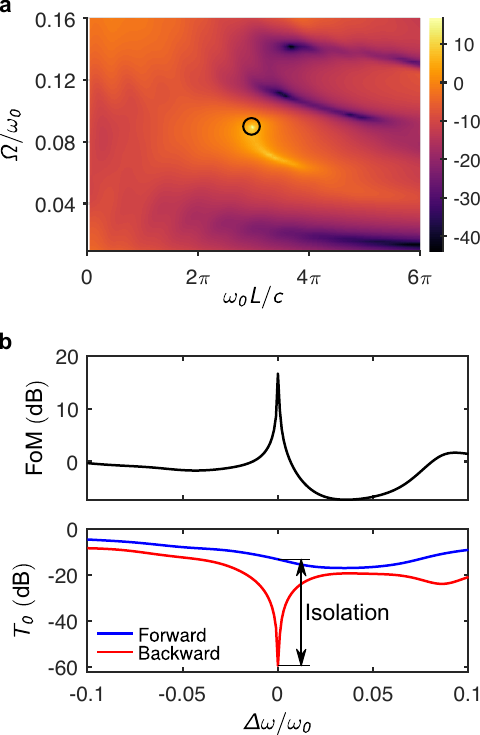}
\caption{Performance analysis of the optical design.
(a) Defined figure of merit (FoM) as a function of the hyperbolic region thickness (normalized to free-space wavelength by multiplication with $k_0=\omega_0/c$) and the modulation frequency (normalized to the optical frequency $\omega_0$). The FoM (color scale) is expressed in dB. The black circle marks the configuration with the maximum FoM.
(b) FoM and zeroth-order transmission as a function of frequency detuning, both in dB scale. The isolation is quantified by $I = 46.4$ dB.}
\label{fig:fig3}
\end{figure}

A practical device must balance isolation with forward insertion loss. To capture this trade-off, we define a figure of merit (FoM) as
\begin{equation}
\text{FoM} = \alpha_1 I + \alpha_2 \, 10 \log_{10}(T_f), \quad \alpha_1 + \alpha_2 = 1,
\label{eq:fom}
\end{equation}
where we assume $\alpha_1=\alpha_2=0.5$ in our analysis. Figure~\hyperref[fig:fig3]{\ref*{fig:fig3}(a)} shows a two-dimensional map of the figure of merit (FoM) as a function of normalized modulation frequency $\Omega/\omega_0$ and hyperbolic thickness $L$. Figure~\hyperref[fig:fig3]{\ref*{fig:fig3}(b)} presents the FoM and the zeroth-order transmission as functions of wavelength detuning, illustrating how performance varies around the design point. Optimal performance is achieved when the modulation frequency is tuned near $\Omega/\omega_0 = 0.1$ and $L \approx 1.5\lambda$, consistent with design constraints in our configuration.

These results highlight the ability of permittivity-modulated slabs to realize strong nonreciprocity at optical frequencies, leveraging reactive modulation to couple harmonics without requiring extreme material parameters, i.e., smaller modulation depths $\Delta \epsilon/\epsilon_{r,0} = 0.1$. The integration of AZO/ZnO hyperbolic multilayers with dynamic dielectric interfaces represents a new paradigm for tunable optical isolators and beam shapers. We note that, unlike the idealized temporal coupled-mode theory (TCMT) scenario of Ref.~\cite{chegnizadeh_non-reciprocity_2020}, our two time-varying slabs are optically thick ($\sqrt{\epsilon_{r,0}}d=3\lambda$) and weakly coupled through a dispersive hyperbolic core. The resulting multi-mode interaction adds internal propagation phases to the effective modulation channels, so that the optimal modulation phase offset for maximum isolation is slightly shifted from the nominal quadrature value. In our design, the FoM peaks at $\phi_2 - \phi_1 \approx \pi/2 + \pi/50$, which we attribute to these higher-order coupling and finite-thickness effects.

\subsection{Microwave Device: 2D Time Crystals}

\begin{figure*}[!t]
\centering
\includegraphics[width=0.99\linewidth]{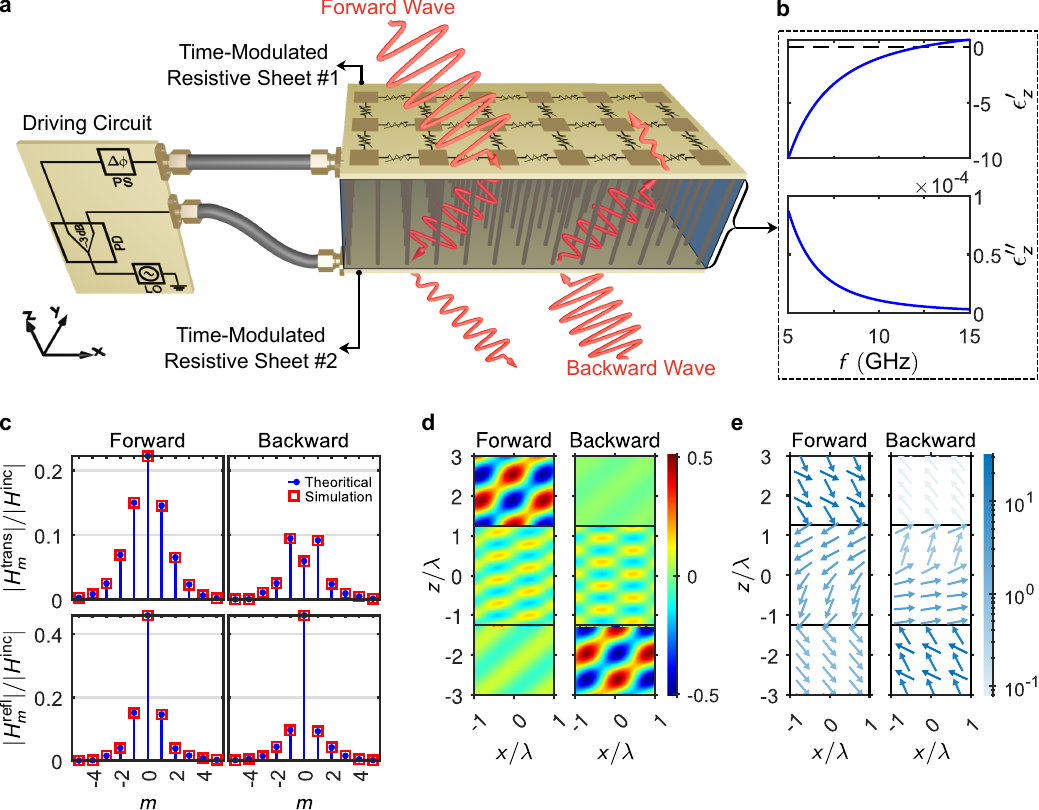}
\caption{Conceptual illustration and representative response of the proposed approach for achieving nonreciprocal negative refraction in the microwave domain. (a) Schematic of the structure made by a wire medium sandwiched by two sheets whose conductance is modulated. Apart from the subwavelength structural parameters of the hyperbolic layer, this system is defined by only two geometrical parameters: hyperbolic medium thickness, $L$, and the incident angle $\theta_{\rm{inc}} = 40^\circ$. The driving circuit, consisting of a source with a modulation frequency of $\Omega$, a power divider (PD), and a phase shifter (PS), provides temporal modulations with different phases. The forward and backward oblique beams exhibit very different transmissions, resulting in nonreciprocity. The red beams (wavy arrows) qualitatively indicate diffraction induced by temporal harmonics and refraction through the structure. (b) Extracted real and imaginary components of the effective transverse permittivity of the wire medium, serving as the hyperbolic layer to enable negative refraction. The lattice constant is $a = 4$ mm along both x and y, with wire radius $r = 0.2$ mm. (c) Simulated and analytically computed amplitudes of the transmitted and reflected fields for both illumination directions, plotted against harmonic index $m$, corresponding to frequency components $\omega_m = \omega_0 + m\Omega$. (d) Spatial profiles of the magnetic field component $H_y$ for forward (top) and backward (bottom) illumination scenarios. Both plots share a common linear color scale shown on the right. (e) Time-averaged Poynting vector fields corresponding to the field maps in panel (d), where arrow direction and color denote energy flow direction and magnitude, respectively. The same logarithmic color scale is used for both cases and displayed on the right.}
\label{fig:fig4}
\end{figure*}

To demonstrate the generality of our framework for achieving time-crystal–enabled nonreciprocal negative refraction, we translate the concept into the microwave regime using conductance-modulated metasurfaces. Here the hyperbolic medium is realized as a wire metamaterial, which is well known to support hyperbolic dispersion at GHz frequencies \cite{simovski_wire_2012}. The interfaces are implemented as resistive sheets whose conductances vary sinusoidally in time, driven by a local oscillator split by a Wilkinson divider and delayed by a 90$^\circ$ phase shifter (Fig.~\hyperref[fig:fig4]{\ref*{fig:fig4}(a)}). This approach closely follows prior designs of time-varying metasurfaces \cite{mostafa_coherently_2022, koutserimpas_multiharmonic_2023}.  

Figure~\hyperref[fig:fig4]{\ref*{fig:fig4}(b)} shows the effective permittivity of the wire medium, confirming hyperbolicity at 10 GHz (derivation in Supplementary). Figures~\hyperref[fig:fig4]{\ref*{fig:fig4}(c)--(e)} present the spectral and spatial results. Harmonic spectra in Fig.~\hyperref[fig:fig4]{\ref*{fig:fig4}(c)} reveal $\sim$11.4 dB isolation between forward and backward zeroth-order transmission. Field maps (Fig.~\hyperref[fig:fig4]{\ref*{fig:fig4}(d)}) show negative refraction in both directions but with much weaker amplitude in the backward case. Poynting vector maps (Fig.~\hyperref[fig:fig4]{\ref*{fig:fig4}(e)}) reinforce this conclusion, visualizing direction-dependent power flow.  

The performance difference compared to the optical device is instructive. While the optical implementation achieved $>40$ dB isolation, the microwave version provides $\sim$11 dB. However, conductance modulation can reach depths as large as $\Delta \sigma/\sigma_0 = 0.9$, far exceeding typical permittivity modulation depths in optics. This ease of achieving strong modulation makes microwave implementations highly practical. Moreover, the sheet-based architecture is significantly simpler than multilayer stacks, lowering fabrication barriers.  

\begin{figure}[!t]
\centering
\includegraphics[width=0.47\linewidth]{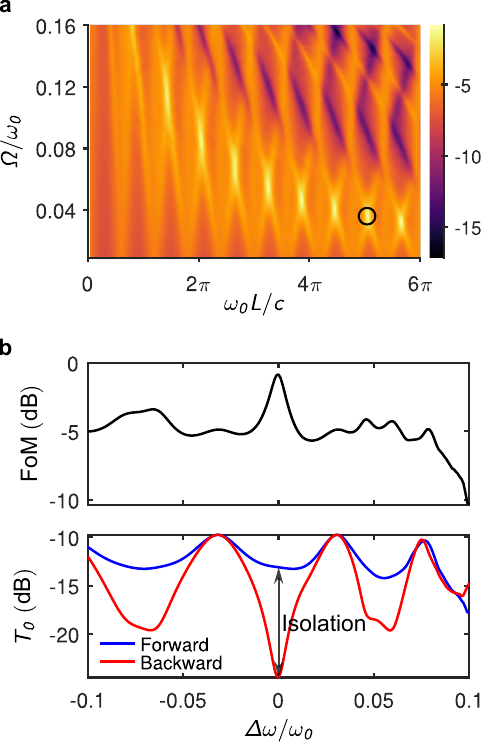}
\caption{Performance analysis of the microwave design.
(a) Figure of merit (FoM) as a function of the hyperbolic region thickness, normalized through $k_0=\omega_0/c$, and the modulation frequency, normalized to the wave frequency $\omega_0$. The FoM, shown on a dB scale, highlights the parameter space leading to the optimal performance.
(b) Frequency detuning dependence of the FoM and zeroth-order transmission, both in dB. The design achieves an isolation of $I = 11.4$ dB, arising from the two time-modulated sheets.}
\label{fig:fig5}
\end{figure}

Figures~\hyperref[fig:fig5]{\ref*{fig:fig5}(a)} and \hyperref[fig:fig5]{\ref*{fig:fig5}(b)} plot the FoM (Eq.~\ref{eq:fom}) across design parameters, identifying operating regimes that maximize isolation while retaining acceptable insertion loss. These maps reveal that even modest detunings can strongly affect performance, emphasizing the importance of careful phase control in the modulation signals.

It is worth studying the microwave device's mechanism and performance in more detail. As mentioned, since the conductive sheets are modulated, the multi-harmonic generation is dependent on the absorptions of the sheets. In fact, when the sheets’ conductivities vary with the modulation frequency, $\Omega$, ohmic losses at different harmonics contribute to one another, collectively producing nonreciprocity at the operating (zeroth-order) frequency, $\omega_0$. To visualize how the total absorption behaves with frequency, we plotted it in the Supplementary.

We have theoretically demonstrated a unified framework for achieving nonreciprocal negative refraction by embedding hyperbolic media between temporally modulated interfaces. The optical implementation, based on permittivity-modulated slabs, leverages reactive modulation to redistribute energy among harmonics and achieves isolation beyond 46 dB. The microwave implementation, based on conductance-modulated metasurfaces, employs absorptive modulation that enforces direction-dependent dissipation, yielding isolation of $\sim$11 dB.  

Together, these results establish the universality of time-crystal–enabled nonreciprocal negative refraction across different frequency regimes and modulation mechanisms. This work demonstrates that temporal modulation can break reciprocity while preserving and shaping negative refraction, thereby expanding the fundamental landscape of photonic time crystals and revealing new possibilities for dynamic control of wave propagation.

\section{Methods}
This section outlines the methodological framework underlying our study. Through compact analytical derivations and harmonic-balance finite element method (FEM) simulations, we capture both the fundamental physics of harmonic generation and the quantitative performance of time-crystal–enabled nonreciprocal negative refraction.

\subsection{Theoretical Analysis}

Our analysis relies on a Floquet-harmonic expansion of the TM-polarized electromagnetic fields, i.e., non-zero components are $E_x$, $E_z$, and $H_y$ ($H_x=H_z=E_y=0$), in configurations with time-varying domains and/or boundaries. This approach is standard for periodically modulated systems, where the modulated areas or elements with frequency of $\Omega$ enforce quasi-energy conservation and produce sidebands at $\omega_m = \omega_0 + m\Omega$ \cite{galiffi_photonics_2022, koutserimpas_multiharmonic_2023}.  

In a time-varying homogeneous region or slab under a TM illumination with the transverse wavevector of $k_{x,{\rm{inc}}}$, the magnetic field takes the multiharmonic form of
\begin{align}
    H^{\mathrm{TV}}_y(x,z;t) &= \sum_m H_{y,m}(x,z) e^{j\omega_m t} \nonumber \\ 
    &= e^{-jk_{x,\mathrm{inc}}x}
\sum_m \left( a_m^{\mathrm{TV}} e^{-jk_{z,m}^{\mathrm{TV}} z} + b_m^{\mathrm{TV}} e^{+jk_{z,m}^{\mathrm{TV}} z} \right) \psi_m^{\mathrm{TV}}(t)
\label{eq:Hfield_TV}
\end{align}
where $a_m^{\mathrm{TV}}$ and $b_m^{\mathrm{TV}}$ are forward and backward modal coefficients, $k_{z,m}^{\mathrm{TV}}$ are the longitudinal propagation constants, and $\psi^{\rm{TV}}_m (t)$ are the multiharmonic eigenmode associated with $k_{z,m}^{\mathrm{TV}}$. It is worth remarking that $\psi_m^{\pm}(t) = e^{\mp jk_{z,m}^{\mathrm{TV}} z}\psi_m^{\mathrm{TV}}(t)$ as previously defined in Conceptual Framework and indicated in Fig. 1(b). In fact, the eigenfunction $\psi^{\rm{TV}}_m (t)$ is a linear combination of harmonics in $\{ \ldots,\allowbreak e^{j\omega_{-2}t},\allowbreak e^{j\omega_{-1}t},\allowbreak e^{j\omega_{0}t},\allowbreak e^{j\omega_{+1}t},\allowbreak e^{j\omega_{+2}t},\allowbreak \ldots \}$, and the $m$th harmonic, i.e., $e^{j\omega_mt}$, is dominant (please see the Supplementary). Thus, each eigenfunction ($m$th mode) for the TV regions would be defined as:
\begin{equation}
\psi^{\rm{TV}}_m (t) = \sum_n \Psi^{\rm{TV}}_{m,n} \, e^{j\omega_n t}
\end{equation}
In the above formula, $\Psi^{\rm{TV}}_{m,n}$ can be derived by dispersion analysis of the time-varying region or 3D time crystals, which depends on the modulation profile and $k_{x,{\rm{inc}}}$ (see the Supplementary for the detailed formulations). The origin of the multiharmonic response here is actually the excitation of such modes. Accordingly, the multiharmonic electric field, i.e., $E^{\mathrm{TV}}_x(x,z;t) = \sum_m E_{x,m}(x,z) e^{j\omega_m t}$, would be obtained from Maxwell’s curl relations by extracting its harmonics ($E_{x,m}$) as:
\begin{equation}
\sum_m \sum_n j\omega_m \, \epsilon_{r,m-n} \, E_{x,n} \, e^{j\omega_m t} = -\frac{\partial H^{\mathrm{TV}}_y(x,z;t)}{\partial z}
\label{eq:Efield_TV}
\end{equation}
where $\epsilon_{r,m-n}$ are the Fourier coefficients of the time-modulated permittivity, i.e., $\epsilon_r (t) = \sum_m \epsilon_{r,m} \, e^{jm\Omega t}$.

This framework allows us to systematically express the fields inside time-varying slabs, where inter-harmonic coupling arises naturally from the Fourier coefficients of the modulation profile. Differently, in static regions, the field expansions remain multi-harmonic, but the harmonics are uncoupled and simply propagate independently.  

For the optical device consisting of two modulated slabs, we employ this formalism to describe the fields in all regions of the structure. Specifically, expansions are written for the upper air region, the upper time-varying (TV) slab, the central hyperbolic medium composed of AZO/ZnO multilayers, the lower TV slab, and the lower air region. The hyperbolic medium itself is treated using effective medium theory (details in the Supplementary), while the TV slabs are modeled using the harmonic expansions in Eqs.~\ref{eq:Hfield_TV} and \ref{eq:Efield_TV}. By enforcing boundary conditions on the tangential field components across all four interfaces, a complete set of coupled equations for the modal coefficients in each region is obtained. A compact block-matrix formulation of this problem is derived and presented in the Supplementary, providing a concrete framework for computing reflection and transmission coefficients for both forward and backward incidence.

A similar procedure is followed for the microwave device, where the time variation is confined to thin resistive metasurfaces rather than volumetric slabs. In this case, the fields in the surrounding static regions (air and wire medium) are expanded in multi-harmonic form, but, as in the optical case, the harmonics are uncoupled in the static domains. The coupling arises solely at the interfaces, where the sheet conductance $\sigma(t)$ is sinusoidally modulated. 

At a sheet with temporally varying conductance $\sigma(t)$, the boundary condition, which is $H^{\mathrm{up}}_y - H^{\mathrm{down}}_y = J_s(t) = \sigma(t) E_x(t)$, couples harmonics via convolution. Expanding $\sigma(t)$ and $E_x(t)$ into Fourier series, we obtain
\begin{equation}
\sum_m \big(H^{\mathrm{up}}_{y,m} - H^{\mathrm{down}}_{y,m}\big) e^{j\omega_m t}
= \sum_m \left(\sum_n \sigma_n E_{x,m-n}\right) e^{j\omega_m t}.
\label{eq:Toeplitz_bc}
\end{equation}
Equation~\ref{eq:Toeplitz_bc} can be written compactly as a Toeplitz matrix acting on the vector of harmonic fields, making explicit how each harmonic at $\omega_m$ couples to its neighbors $\omega_{m\pm 1}$ through the Fourier coefficients $\sigma_n$ of the modulation profile.  

Accordingly, the full system for the microwave implementation consists of three regions separated by two time-modulated boundaries: the upper air domain, the upper time-modulated sheet, the hyperbolic wire medium, the lower time-modulated sheet, and the lower air domain. The wire medium itself is homogenized using a Drude-type effective permittivity along the wire axis, while supporting hyperbolic dispersion in the transverse plane \cite{simovski_wire_2012}. As before, continuity of tangential electric fields across the time-varying sheets is enforced, but for the magnetic fields, surface-current terms induce multiharmonic discontinuities. This leads to a coupled set of equations relating reflection, transmission, and internal modal coefficients across all harmonics. A compact block-matrix formulation of this system is also provided in the Supplementary, in direct analogy to the slab-based optical case.

\subsection{Numerical Simulation}
To verify our theoretical analysis, we employed full-wave simulations. Typical FEM solvers that compute the electromagnetic response in the frequency domain do not inherently support the simulation of time-varying configurations, and additional setups are required to simulate a time-varying structure accurately. Since our modulation scheme is periodic and even harmonic (sinusoidal), we expected to observe multi-harmonic frequency responses in every region. Thus, for such a scenario, it is more efficient to simulate the problem in the frequency domain rather than the time domain, e.g., the finite difference time domain (FDTD) method. Therefore, we have chosen FEM solvers. It is evident that FEM solvers can simulate geometrically complex time-varying problems more effectively than other frequency-domain methods, such as the finite difference frequency domain (FDFD) \cite{shi_multi-frequency_2016}, making this type of solver a reasonable choice for customization and development, thanks to the conformal meshing capabilities in FEM. Additionally, unlike previous studies where numerical simulations and postprocessing relied on slow (adiabatic) temporal modulations, the multi-harmonic coupled FEM solvers used here can accurately capture the system’s response under fast modulation conditions \cite{salary_electrically_2018, sedeh_optical_2022}. Consequently, this simulation framework eliminates the need for temporal adiabatic assumptions, ensuring correct and stable results even when the modulation frequency is high.

As mentioned previously, we must implement additional setups in a typical or commercial FEM solver to carry out a full-wave simulation for time-varying (time-modulated) configurations. In this fashion, we should consider an FEM solver for each of the harmonics we are interested in to find the response. The pivotal step is to couple all of these FEM solvers together (Harmonic-Balance FEM) based on the modulation profile and the time-varying regions. To be more specific, for the first optical design, we had time-modulated slabs (3D time crystals), which are in fact regions or domains whose permittivity is modulated. In this setup, the time modulations induce volumetric polarizations at adjacent harmonics of each harmonic $\omega_m$, i.e.,  $\omega_m \pm \Omega$, whose amplitude and phase are determined based on the modulation profile of that region. Likewise, for the second microwave configuration, including time-modulated surfaces or sheets (2D time crystals), the surface conductance of the sheets is modulated. In this case, the temporal variation of the sheet conductance generates surface currents at neighboring harmonics, shifted by $\pm\Omega$ from each $\omega_m$, with their strengths determined by the modulation waveform imposed on the interface. The coupled FEM solvers in both configurations (optical and microwave designs) can compute the multi-harmonic response. In such a tailored simulation setup, the computational dimension of the full-wave simulation is folded by the number of harmonics studied through the coupled solvers. Importantly, our simulation results show excellent agreement with the analytical approach described earlier. In particular, the harmonic spectra plotted in Figs.~\hyperref[fig:fig2]{\ref*{fig:fig2}(c)} and ~\hyperref[fig:fig4]{\ref*{fig:fig4}(c)} show perfect agreement between theory and simulation for the optical and microwave devices, respectively, validating the accuracy of both methods.

\section*{Data availability}
Additional data are available from the corresponding author upon reasonable request.

\section*{Acknowledgements}
This work was supported in part by the National Science Foundation (NSF) under Grant No. DMR-2323909.

\section*{Author contributions}
M.R.T. conceived the idea, developed the theoretical framework, and performed the simulations and analysis. W.C. supervised the project and provided conceptual guidance. Both authors discussed the results and jointly wrote and revised the manuscript.

\section*{Competing interests}
The authors declare no competing interests. 

\section*{Additional information}
Supplementary information is available for this paper. Correspondence and requests for materials should be addressed to W.C.


\bibliographystyle{naturemag}  
\bibliography{Refs}

\end{document}


\maketitle

\tableofcontents
\clearpage


\section{Optical configuration}
\subsection{Matrix formulation for time-modulated-media dispersion}

We derive here the dispersion relation for transverse–magnetic (TM) waves in a dielectric whose
relative permittivity varies periodically in time. Starting from Maxwell's curl equations,
\[
\nabla \times \mathbf{E}(t) = -\mu_0 \frac{\partial \mathbf{H}(t)}{\partial t},
\qquad
\nabla \times \mathbf{H}(t) = \frac{\partial}{\partial t}\!\left[\epsilon_0\,\epsilon_r(t)\mathbf{E}(t)\right],
\]
and eliminating $\mathbf{E}(t)$, we obtain the magnetic–field wave equation
\begin{equation}
\nabla^2 \mathbf{H}(t)
-\frac{1}{c^2}\frac{\partial}{\partial t}
\left[\epsilon_r(t)\frac{\partial \mathbf{H}(t)}{\partial t}\right]
= 0.
\label{eq:wave-time}
\end{equation}

We focus on TM polarization and assume spatial invariance along $y$, so that
\[
\mathbf{H}(x,z,t)=H_y(x,z,t)\,\hat{\mathbf{y}}.
\]
The temporally periodic relative permittivity is expanded as
\[
\epsilon_r(t)=\sum_{n\in\mathbb{Z}}\epsilon_{r,n}\,e^{jn\Omega t},
\qquad
T=\frac{2\pi}{\Omega},
\]
and the magnetic field admits the Floquet expansion
\[
H_y(x,z,t)=e^{-jk_x x}\sum_{m\in\mathbb{Z}}H_m(z)\,e^{j(\omega_0+m\Omega)t}.
\]

Substituting these series into Eq.~\eqref{eq:wave-time} and matching coefficients of
$e^{j(\omega_0+m\Omega)t}$ gives the coupled-harmonic relation
\begin{equation}
\bigl(k_x^2 + k_{z,m}^2\bigr) H_m
=
\frac{1}{c^2}
\sum_{n\in\mathbb{Z}}
\epsilon_{r,n}\,
(\omega_0+m\Omega)(\omega_0+(m-n)\Omega)\,
H_{m-n}.
\label{eq:harmonic-coupling}
\end{equation}
where $k_{z,m}$ is the longitudinal wavenumber of the $m$th harmonic.

\paragraph{Compact matrix representation.}
We define the multiharmonic amplitude vector
\[
\underline{H}=(\ldots,H_{-1},H_0,H_{+1},\ldots)^{\mathsf{T}},
\]
the diagonal operator
\[
\underline{\underline{k}}_z^2=\mathrm{diag}(\ldots,k_{z,-1}^2,k_{z,0}^2,k_{z,+1}^2,\ldots),
\]
and the diagonal frequency operator
\[
\underline{\underline{\omega}}
=\mathrm{diag}(\ldots,\omega_{-1},\omega_0,\omega_{+1},\ldots),
\qquad
\omega_m=\omega_0+m\Omega.
\]
The Toeplitz permittivity operator is written as
\[
\underline{\underline{\epsilon_r}}=
\begin{pmatrix}
\ddots & \vdots & \vdots & \vdots & \reflectbox{$\ddots$} \\
\cdots & \epsilon_{r,0} & \epsilon_{r,-1} & \epsilon_{r,-2} & \cdots \\
\cdots & \epsilon_{r,+1} & \epsilon_{r,0} & \epsilon_{r,-1} & \cdots \\
\cdots & \epsilon_{r,+2} & \epsilon_{r,+1} & \epsilon_{r,0} & \cdots \\
\reflectbox{$\ddots$} & \vdots & \vdots & \vdots & \ddots
\end{pmatrix}.
\]

Equation~\eqref{eq:harmonic-coupling} becomes
\[
\left(k_x^2\,\underline{\underline{I}}+\underline{\underline{k}}_z^2\right)\underline{H}
=
\frac{1}{c^2}\,
\underline{\underline{\omega}}\,
\underline{\underline{\epsilon_r}}\,
\underline{\underline{\omega}}\,
\underline{H},
\]
which yields the generalized eigenvalue problem
\begin{equation}
\Bigl[
k_x^2\,\underline{\underline{I}}
+\underline{\underline{k}}_z^2
-\frac{1}{c^2}\,
\underline{\underline{\omega}}\,
\underline{\underline{\epsilon_r}}\,
\underline{\underline{\omega}}
\Bigr]
\underline{H}
=0.
\label{eq:generalized-eig}
\end{equation}
Nontrivial solutions require the dispersion condition
\begin{equation}
\det\!\left(
k_x^2\,\underline{\underline{I}}
+\underline{\underline{k}}_z^2
-\frac{1}{c^2}\,
\underline{\underline{\omega}}\,
\underline{\underline{\epsilon_r}}\,
\underline{\underline{\omega}}
\right)=0.
\label{eq:dispersion}
\end{equation}

\paragraph{Electric field reconstruction.}
Once the eigenpair $(k_{z,m},\underline{H})$ is determined, the harmonic electric-field coefficients follow from Maxwell’s equations:
\[
E_m(x,z)=
\epsilon_0^{-1}\,
\underline{\underline{\epsilon_r}}^{-1}
\left(
\frac{k_{z,m}}{\omega_m}\hat{\mathbf{x}}
-
\frac{k_x}{\omega_m}\hat{\mathbf{z}}
\right)
H_m,
\]
and the total field is reconstructed as
\[
\mathbf{E}(x,z,t)=\sum_{m\in\mathbb{Z}}E_m(x,z)\,e^{j(\omega_0+m\Omega)t}.
\]

\begin{figure*}[!t]
\centering
\includegraphics[width=0.95\linewidth]{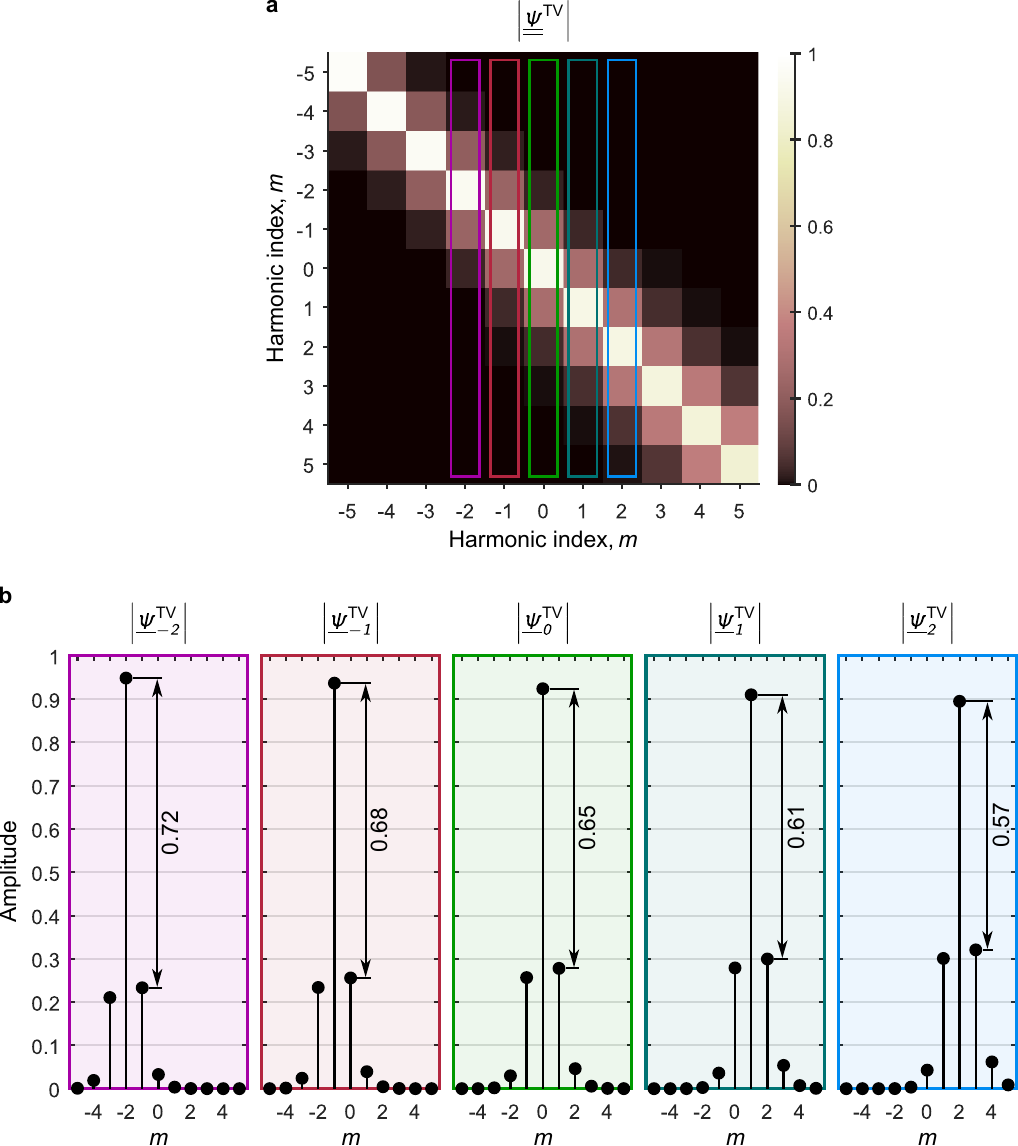}
\caption{Multiharmonic eigenmodes of the two time-varying (TV) slabs.
(a) Amplitude map of the eigenfunction matrices 
$\underline{\underline{\Psi}}^{\mathrm{TV1}}$ and $\underline{\underline{\Psi}}^{\mathrm{TV2}}$, 
whose columns correspond to the dominant multiharmonic modes supported by each temporally 
modulated slab. The colored rectangles highlight the five principal modes, 
$\underline{\Psi}^{\mathrm{TV}}_{-2}$, $\underline{\Psi}^{\mathrm{TV}}_{-1}$, 
$\underline{\Psi}^{\mathrm{TV}}_{0}$, $\underline{\Psi}^{\mathrm{TV}}_{+1}$, 
and $\underline{\Psi}^{\mathrm{TV}}_{+2}$, each of which comprises a mixture of harmonics indexed by $m$. (b) Stem plots of the harmonic amplitudes for the five modes identified in (a). For each multiharmonic mode, the relative weights of the primary harmonic and its side harmonics are shown, illustrating how the temporal modulation redistributes energy across adjacent frequency components.  Each stem-plot panel is color-matched to its corresponding column in (a). }
\label{fig:figS1}
\end{figure*}

Figure~\ref{fig:figS1} illustrates representative multiharmonic eigenmodes obtained from the dispersion
equation~\eqref{eq:generalized-eig}. Each eigenmode of the time-varying slab is represented by the
column vector $\underline{\Psi}^{\mathrm{TV}}_{m}$, whose elements give the complex amplitudes of the
harmonics $\{\ldots,\,e^{j\omega_{-1}t},\,e^{j\omega_0 t},\,e^{j\omega_{+1}t},\ldots\}$ that collectively form the
$m$th Floquet eigenmode. The full eigenfunction matrix
\[
\underline{\underline{\Psi}}^{\mathrm{TV}}
=\bigl[\ldots,\underline{\Psi}^{\mathrm{TV}}_{-2},
\underline{\Psi}^{\mathrm{TV}}_{-1},
\underline{\Psi}^{\mathrm{TV}}_{0},
\underline{\Psi}^{\mathrm{TV}}_{+1},
\underline{\Psi}^{\mathrm{TV}}_{+2},\ldots\bigr]
\]
is therefore constructed by stacking these modal vectors as its columns. As seen in
Fig.~\hyperref[fig:figS1]{\ref*{fig:figS1}(b)}, the relative level of the side harmonics grows progressively as we move from the mode
with index $m=-2$ toward $m=+2$, indicating that the temporal modulation induces slightly stronger harmonic
mixing for the positive-index modes. This trend reflects the asymmetric coupling among
Floquet orders introduced by the time modulation and quantitatively captured by the matrix
formulation above.

\subsection{Analysis of the multilayer structure composed of two modulated slabs surrounding a hyperbolic medium}
\label{sec:multilayer-analysis}

We now specialize the multiharmonic formalism to the multilayer system used in the main text,
consisting of two time-modulated dielectric slabs surrounding a static hyperbolic medium. The
magnetic field in each region is expanded as
\begin{align*}
\mathbf{H}^{\rm I}(x,z;t)
&=
\Bigl[
e^{-j k_{x,{\rm inc}} x}\,e^{-j k_{z,{\rm inc}} z}\,e^{j\omega_0 t}
+
e^{-j k_{x,{\rm inc}} x}
\sum_m r_m\,e^{+j k_{z,m} z} e^{j\omega_m t}
\Bigr]\hat{\mathbf{y}}, \\
\mathbf{H}^{\rm TV1}(x,z;t)
&=
e^{-j k_{x,{\rm inc}} x}
\sum_m
\Bigl(
a^{\rm TV1}_m e^{-j k^{\rm TV1}_{z,m} z}
+
b^{\rm TV1}_m e^{+j k^{\rm TV1}_{z,m} z}
\Bigr)
\psi^{\rm TV1}_m(t)\,\hat{\mathbf{y}}, \\
\mathbf{H}^{\rm II}(x,z;t)
&=
e^{-j k_{x,{\rm inc}} x}
\sum_m
\Bigl(
a_m e^{-j k^{\rm hyp}_{z,m} z}
+
b_m e^{+j k^{\rm hyp}_{z,m} z}
\Bigr)e^{j\omega_m t}\,\hat{\mathbf{y}}, \\
\mathbf{H}^{\rm TV2}(x,z;t)
&=
e^{-j k_{x,{\rm inc}} x}
\sum_m
\Bigl(
a^{\rm TV2}_m e^{-j k^{\rm TV2}_{z,m} z}
+
b^{\rm TV2}_m e^{+j k^{\rm TV2}_{z,m} z}
\Bigr)
\psi^{\rm TV2}_m(t)\,\hat{\mathbf{y}}, \\
\mathbf{H}^{\rm III}(x,z;t)
&=
e^{-j k_{x,{\rm inc}} x}
\sum_m
t_m e^{-j k_{z,m} z} e^{j\omega_m t}\,\hat{\mathbf{y}},
\end{align*}
where $\omega_m=\omega_0+m\Omega$, and $\psi^{\rm TV1}_m(t)$ and $\psi^{\rm TV2}_m(t)$ are the
eigenfunctions of the time-varying slabs in regions TV1 and TV2, respectively, associated with the
eigenvalues $k^{\rm TV1}_{z,m}$ and $k^{\rm TV2}_{z,m}$ (see Sec.~1.1).

\paragraph{Layer geometry.}
The $z$–stratified architecture is
\begin{align*}
z<-d-\tfrac{L}{2} &: \text{air (incidence region I)}, \\
-d-\tfrac{L}{2}<z<-\tfrac{L}{2} &: \text{time-varying slab 1 (TV1)}, \\
-\tfrac{L}{2}<z<\tfrac{L}{2} &: \text{static hyperbolic slab (region II)}, \\
\tfrac{L}{2}<z<\tfrac{L}{2}+d &: \text{time-varying slab 2 (TV2)}, \\
\tfrac{L}{2}+d<z &: \text{air (transmission region III)},
\end{align*}
where $d$ is the thickness of each modulated slab and $L$ is the thickness of the hyperbolic core.

\paragraph{Multiharmonic vectors and operators.}
We collect the harmonic amplitudes into column vectors,
\begin{align*}
&\underline{r}
\triangleq (\ldots,r_{-2},r_{-1},r_0,r_{+1},r_{+2},\ldots)^{\mathsf{T}},\quad
\underline{a^{\rm TV1}}
\triangleq (\ldots,a^{\rm TV1}_{-2},a^{\rm TV1}_{-1},a^{\rm TV1}_0,a^{\rm TV1}_{+1},\ldots)^{\mathsf{T}}, \\
&\underline{b^{\rm TV1}},\;
\underline{a},\;
\underline{b},\;
\underline{a^{\rm TV2}},\;
\underline{b^{\rm TV2}},\;
\underline{t}
\quad\text{defined analogously,}
\end{align*}
and define the diagonal operators
\begin{align*}
\underline{\underline{k_z}}
&\triangleq \mathrm{diag}(\ldots,k_{z,-2},k_{z,-1},k_{z,0},k_{z,+1},k_{z,+2},\ldots),
\quad k_{z,m} = \sqrt{\Bigl(\tfrac{\omega_m}{c}\Bigr)^2-k^2_{x,{\rm inc}}}, \\
\underline{\underline{k^{\rm hyp}_z}}
&\triangleq \mathrm{diag}(\ldots,k^{\rm hyp}_{z,-2},k^{\rm hyp}_{z,-1},\ldots),
\quad k^{\rm hyp}_{z,m} = \sqrt{\epsilon_t\Bigl[\Bigl(\tfrac{\omega_m}{c}\Bigr)^2-\tfrac{k_{x,{\rm inc}}^2}{\epsilon_z}\Bigr]}, \\
\underline{\underline{k^{\rm TV1}_z}},
\quad
&\underline{\underline{k^{\rm TV2}_z}}
\quad\text{(diagonal matrices of TV1/TV2 longitudinal wavenumbers)}, \\
\underline{\underline{\omega}}
&\triangleq \mathrm{diag}(\ldots,\omega_{-2},\omega_{-1},\omega_0,\omega_{+1},\omega_{+2},\ldots),
\quad \omega_m =\omega_0+m\Omega.
\end{align*}

The temporally modulated permittivity profiles in TV1 and TV2 are represented by Toeplitz
operators
\[
\underline{\underline{\epsilon_r^{\rm TV1}}}
\quad\text{and}\quad
\underline{\underline{\epsilon_r^{\rm TV2}}},
\]
whose $(p,q)$ elements are given by the corresponding permittivity harmonic
$\epsilon^{\rm TV1}_{r,p-q}$ or $\epsilon^{\rm TV2}_{r,p-q}$, respectively.
We also define the unit vector
\[
\underline{1_0}
\triangleq (\ldots,0,0,1,0,0,\ldots)^{\mathsf{T}},
\]
with a single nonzero entry at the central (fundamental) harmonic.

\paragraph{Modal matrices of the time-varying slabs.}
For each time-varying region, the $m$th multiharmonic eigenmode is represented by the column
vector $\underline{\Psi}^{\rm TV1}_m$ or $\underline{\Psi}^{\rm TV2}_m$, whose entries are the
complex amplitudes of the harmonics
$\{\ldots,e^{j\omega_{-2}t},e^{j\omega_{-1}t},e^{j\omega_0 t},e^{j\omega_{+1}t},\ldots\}$ in the
eigenfunction $\psi^{\rm TV1}_m(t)$ or $\psi^{\rm TV2}_m(t)$. We collect these into the modal
matrices
\[
\underline{\underline{\Psi}}^{\rm TV1}
\triangleq (\ldots,\underline{\Psi}^{\rm TV1}_{-2},\underline{\Psi}^{\rm TV1}_{-1},
\underline{\Psi}^{\rm TV1}_0,\underline{\Psi}^{\rm TV1}_{+1},\underline{\Psi}^{\rm TV1}_{+2},\ldots),
\quad
\underline{\underline{\Psi}}^{\rm TV2}
\triangleq (\ldots,\underline{\Psi}^{\rm TV2}_{-2},\ldots,\underline{\Psi}^{\rm TV2}_{+2},\ldots),
\]
so that columns index the eigenmodes and rows index the temporal harmonics, consistent with the
notation introduced in Fig.~\ref{fig:figS1}.

\paragraph{Multiharmonic magnetic fields in each region.}
Using the above definitions, the harmonic-domain magnetic fields in the five regions can be
written compactly as
\begin{equation}
\begin{aligned}
\underline{\mathbf{H}}^{\rm I}(x,z)
&=
e^{-j k_{x,{\rm inc}} x}
\Bigl[
e^{-j k_{z,{\rm inc}} z}\,\underline{1_0}
+
e^{+j\underline{\underline{k_z}} z}\,\underline{r}
\Bigr]\hat{\mathbf{y}},\\[2pt]
\underline{\mathbf{H}}^{\rm TV1}(x,z)
&=
e^{-j k_{x,{\rm inc}} x}\,
\underline{\underline{\Psi}}^{\rm TV1}
\Bigl[
e^{-j\underline{\underline{k^{\rm TV1}_z}} z}\,\underline{a^{\rm TV1}}
+
e^{+j\underline{\underline{k^{\rm TV1}_z}} z}\,\underline{b^{\rm TV1}}
\Bigr]\hat{\mathbf{y}},\\[2pt]
\underline{\mathbf{H}}^{\rm II}(x,z)
&=
e^{-j k_{x,{\rm inc}} x}
\Bigl[
e^{-j\underline{\underline{k^{\rm hyp}_z}} z}\,\underline{a}
+
e^{+j\underline{\underline{k^{\rm hyp}_z}} z}\,\underline{b}
\Bigr]\hat{\mathbf{y}},\\[2pt]
\underline{\mathbf{H}}^{\rm TV2}(x,z)
&=
e^{-j k_{x,{\rm inc}} x}\,
\underline{\underline{\Psi}}^{\rm TV2}
\Bigl[
e^{-j\underline{\underline{k^{\rm TV2}_z}} z}\,\underline{a^{\rm TV2}}
+
e^{+j\underline{\underline{k^{\rm TV2}_z}} z}\,\underline{b^{\rm TV2}}
\Bigr]\hat{\mathbf{y}},\\[2pt]
\underline{\mathbf{H}}^{\rm III}(x,z)
&=
e^{-j k_{x,{\rm inc}} x}\,
e^{-j\underline{\underline{k_z}} z}\,\underline{t}\,\hat{\mathbf{y}}.
\end{aligned}
\label{eq:H-matrix-regions}
\end{equation}

\paragraph{Electric fields from Maxwell’s equations.}
We similarly collect the harmonic amplitudes of the electric-field components into
$\underline{\mathbf{E}}_x$ and $\underline{\mathbf{E}}_z$. In the time-varying slabs, the
multiharmonic constitutive relation yields
\begin{align}
\underline{\mathbf{E}}_x^{\rm TV1,TV2}(x,z)
&=
-\frac{1}{j\epsilon_0}
\Bigl(
\underline{\underline{\omega}}\,
\underline{\underline{\epsilon_r^{\rm TV1,TV2}}}
\Bigr)^{-1}
\frac{\partial\underline{\mathbf{H}}^{\rm TV1,TV2}(x,z)}{\partial z},
\notag\\[2pt]
\underline{\mathbf{E}}_z^{\rm TV1,TV2}(x,z)
&=
\frac{1}{j\epsilon_0}
\Bigl(
\underline{\underline{\omega}}\,
\underline{\underline{\epsilon_r^{\rm TV1,TV2}}}
\Bigr)^{-1}
\frac{\partial\underline{\mathbf{H}}^{\rm TV1,TV2}(x,z)}{\partial x},
\label{eq:ExEz-TV}
\end{align}
where the differential operators act element-wise on the $z$-dependent factors in
Eq.~\eqref{eq:H-matrix-regions}.

For the static regions, it is convenient to introduce the diagonal matrix of total wavenumbers
\[
\underline{\underline{k}}
\triangleq \mathrm{diag}(\ldots,k_{-2},k_{-1},k_0,k_{+1},k_{+2},\ldots),
\qquad
k_m=\frac{\omega_m}{c},
\]
and the diagonal propagation-angle matrix
\[
\underline{\underline{\theta}}
\triangleq \mathrm{diag}(\ldots,\theta_{-2},\theta_{-1},\theta_0,\theta_{+1},\theta_{+2},\ldots),
\qquad
\theta_m=\cos^{-1}\!\Bigl(\frac{k_{z,m}}{k_m}\Bigr),
\]
together with their element-wise cosine and sine matrices,
\[
\cos\underline{\underline{\theta}}
\triangleq \mathrm{diag}(\ldots,\cos\theta_{-2},\ldots,\cos\theta_{+2},\ldots),
\quad
\sin\underline{\underline{\theta}}
\triangleq \mathrm{diag}(\ldots,\sin\theta_{-2},\ldots,\sin\theta_{+2},\ldots).
\]

Using the usual TM relations, the $x$–component of the electric field in each region can then be
expressed as
\begin{align*}
\underline{\mathbf{E}}_x^{\rm I}(x,z)
&=
\eta\,e^{-j k_{x,{\rm inc}} x}
\Bigl[
\cos\theta_{\rm inc}\,e^{-j k_{z,{\rm inc}} z}\,\underline{1_0}
-
\cos\underline{\underline{\theta}}\,
e^{+j\underline{\underline{k_z}} z}\,\underline{r}
\Bigr],\\[2pt]
\underline{\mathbf{E}}_x^{\rm TV1}(x,z)
&=
\epsilon_0^{-1}\,
e^{-j k_{x,{\rm inc}} x}
\Bigl(
\underline{\underline{\omega}}\,
\underline{\underline{\epsilon_r^{\rm TV1}}}
\Bigr)^{-1}
\underline{\underline{\Psi}}^{\rm TV1}\,
\underline{\underline{k_z^{\rm TV1}}}
\Bigl[
e^{-j\underline{\underline{k_z^{\rm TV1}}} z}\,\underline{a^{\rm TV1}}
-
e^{+j\underline{\underline{k_z^{\rm TV1}}} z}\,\underline{b^{\rm TV1}}
\Bigr],\\[2pt]
\underline{\mathbf{E}}_x^{\rm II}(x,z)
&=
\eta\,e^{-j k_{x,{\rm inc}} x}\,
\underline{\underline{k_z^{\rm hyp}}}\,
\underline{\underline{k}}^{-1}
\Bigl[
e^{-j\underline{\underline{k_z^{\rm hyp}}} z}\,\underline{a}
-
e^{+j\underline{\underline{k_z^{\rm hyp}}} z}\,\underline{b}
\Bigr]\frac{1}{\epsilon_t},\\[2pt]
\underline{\mathbf{E}}_x^{\rm TV2}(x,z)
&=
\epsilon_0^{-1}\,
e^{-j k_{x,{\rm inc}} x}
\Bigl(
\underline{\underline{\omega}}\,
\underline{\underline{\epsilon_r^{\rm TV2}}}
\Bigr)^{-1}
\underline{\underline{\Psi}}^{\rm TV2}\,
\underline{\underline{k_z^{\rm TV2}}}
\Bigl[
e^{-j\underline{\underline{k_z^{\rm TV2}}} z}\,\underline{a^{\rm TV2}}
-
e^{+j\underline{\underline{k_z^{\rm TV2}}} z}\,\underline{b^{\rm TV2}}
\Bigr],\\[2pt]
\underline{\mathbf{E}}_x^{\rm III}(x,z)
&=
\eta\,e^{-j k_{x,{\rm inc}} x}\,
\cos\underline{\underline{\theta}}\,
e^{-j\underline{\underline{k_z}} z}\,\underline{t}.
\end{align*}

\subsubsection*{Boundary conditions at the interfaces}

The four interfaces are located at
\[
z_1 = -d-\frac{L}{2},\qquad
z_2 = -\frac{L}{2},\qquad
z_3 = \frac{L}{2},\qquad
z_4 = \frac{L}{2}+d.
\]
At each interface, we enforce continuity of the tangential fields $H_y$ and $E_x$. Evaluating
Eqs.~\eqref{eq:H-matrix-regions} and the above expressions for $\underline{\mathbf{E}}_x$ at
$z=z_i$ leads to a set of linear relations among the unknown coefficient vectors
$\underline{r}$, $\underline{a^{\rm TV1}}$, $\underline{b^{\rm TV1}}$,
$\underline{a}$, $\underline{b}$, $\underline{a^{\rm TV2}}$, $\underline{b^{\rm TV2}}$,
and $\underline{t}$. For example, at the first interface $z=z_1$ (between regions I and TV1) we
obtain
\begin{align*}
e^{-j k_{z,{\rm inc}} z_1}\,\underline{1_0}
+
e^{+j\underline{\underline{k_z}} z_1}\,\underline{r}
&=
\underline{\underline{\Psi}}^{\rm TV1}
\Bigl[
e^{-j\underline{\underline{k_z^{\rm TV1}}} z_1}\,\underline{a^{\rm TV1}}
+
e^{+j\underline{\underline{k_z^{\rm TV1}}} z_1}\,\underline{b^{\rm TV1}}
\Bigr],\\[2pt]
\eta\Bigl[
\cos\theta_{\rm inc}\,e^{-j k_{z,{\rm inc}} z_1}\,\underline{1_0}
-
\cos\underline{\underline{\theta}}\,
e^{+j\underline{\underline{k_z}} z_1}\,\underline{r}
\Bigr]
&=
\epsilon_0^{-1}
\Bigl(
\underline{\underline{\omega}}\,
\underline{\underline{\epsilon_r^{\rm TV1}}}
\Bigr)^{-1}
\underline{\underline{\Psi}}^{\rm TV1}\,
\underline{\underline{k_z^{\rm TV1}}}
\Bigl[
e^{-j\underline{\underline{k_z^{\rm TV1}}} z_1}\,\underline{a^{\rm TV1}}
-
e^{+j\underline{\underline{k_z^{\rm TV1}}} z_1}\,\underline{b^{\rm TV1}}
\Bigr].
\end{align*}
Analogous pairs of continuity equations are obtained at $z=z_2$ (TV1–hyperbolic),
$z=z_3$ (hyperbolic–TV2), and $z=z_4$ (TV2–air), by substituting the corresponding field
expressions in regions II, TV2, and III. Collectively, these boundary conditions form a
multiharmonic linear system for the unknown vectors
$\underline{r}$, $\underline{a^{\rm TV1}}$, $\underline{b^{\rm TV1}}$,
$\underline{a}$, $\underline{b}$, $\underline{a^{\rm TV2}}$, $\underline{b^{\rm TV2}}$,
and $\underline{t}$, which is solved numerically in our simulations.

\subsection{Performance analysis of the optical device in more detail}

\begin{figure*}[!t]
\centering
\includegraphics[width=0.76\linewidth]{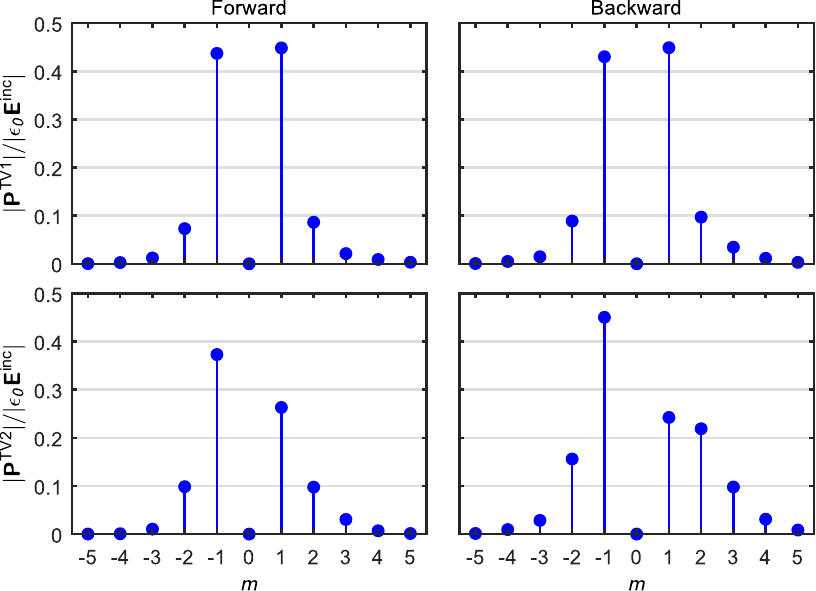}
\caption{Harmonic polarization amplitudes induced in the two time-varying slabs. Polarization amplitudes of the generated Floquet harmonics in the two temporally modulated slabs, computed under forward and backward illumination. Each stem plot shows the magnitude of the volumetric polarization at harmonic index $m$, normalized to the incident flux density. The distributions highlight how temporal modulation excites distinct sets of harmonics in the forward and backward directions, reflecting the direction-dependent coupling that underlies the nonreciprocal response.}
\label{fig:figS2}
\end{figure*}

Figure~\ref{fig:figS2} provides additional insight into the multiharmonic polarization generated inside the time-varying dielectric slabs. Because the relative permittivity is periodically modulated,
\[
\epsilon_r(t)=\sum_{n}\epsilon_{r,n} e^{jn\Omega t}
\quad\xleftrightarrow{\;\mathcal{F}\;}\quad
\epsilon_r(\omega)
=
2\pi\sum_{n}\epsilon_{r,n}\delta(\omega-n\Omega),
\]
the material response consists of discrete spectral lines spaced by the modulation frequency. The induced polarization follows
\[
\mathbf{P}(t)=\epsilon_0\,\epsilon_r(t)\mathbf{E}(t)
\quad\xleftrightarrow{\;\mathcal{F}\;}\quad
\mathbf{P}(\omega)
=
2\pi\epsilon_0\sum_{n}\epsilon_{r,n}\mathbf{E}(\omega-n\Omega),
\]
which reflects the multiply–convolve property of Fourier transforms. Writing the fields in the Floquet form,
\[
\mathbf{E}(t)=\sum_{m}\mathbf{E}_m e^{jm\Omega t},
\qquad
\mathbf{P}(t)=\sum_{m}\mathbf{P}_m e^{jm\Omega t},
\]
gives
\[
\mathbf{P}_m=\epsilon_0\sum_{n}\epsilon_{r,n}\mathbf{E}_{m-n}.
\]
This relation explains the trend shown in Fig.~\ref{fig:figS2}: the modulation redistributes energy among neighboring sidebands, and the relative gap between the principal and side harmonics slowly decreases as we move from lower-order to higher-order eigenmodes, consistent with the modal structure illustrated in Fig.~\hyperref[fig:figS1]{\ref*{fig:figS1}(b)} previously.

\begin{figure}[!t]
\centering
\includegraphics[width=0.99\linewidth]{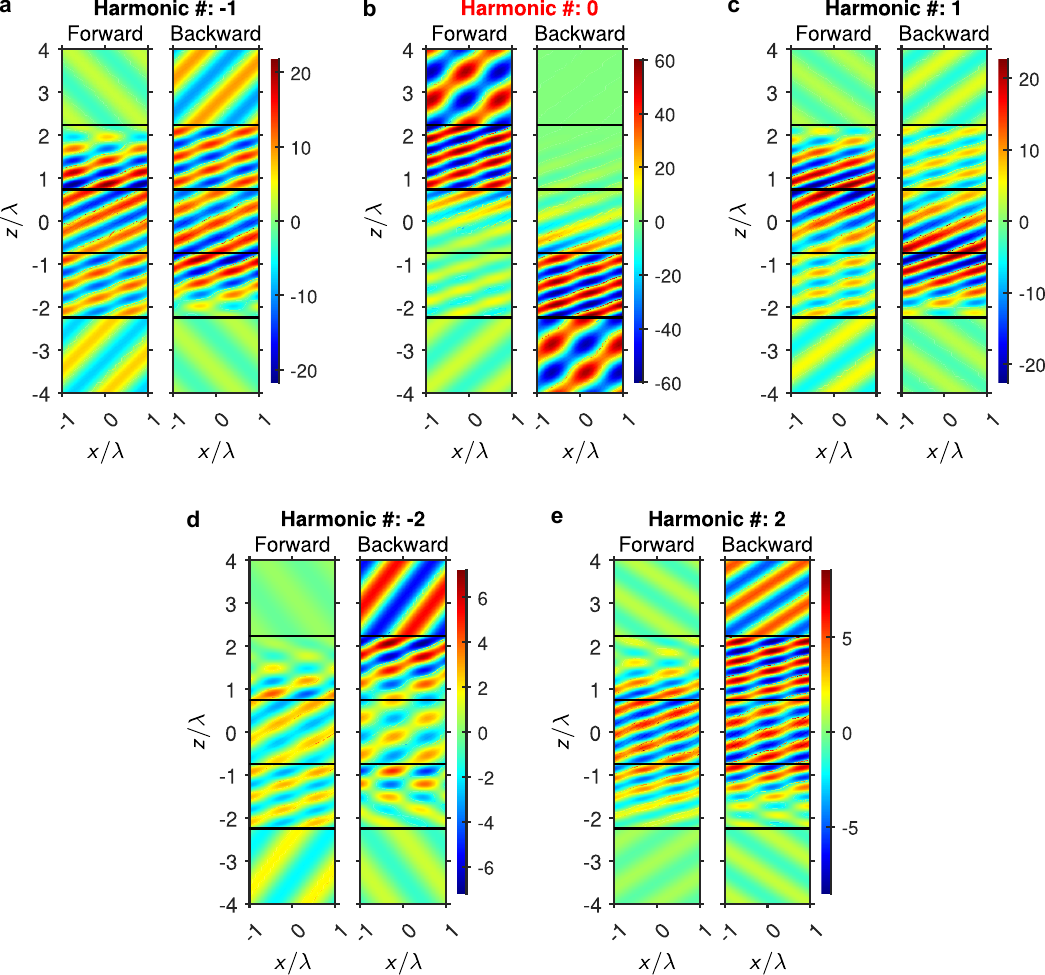}
\caption{Magnetic-field distributions at the harmonic frequencies 
$\omega_{m} = \omega_{0} + m\Omega$ for the optical device. Spatial distributions of the magnetic field component $H_{y}(x,z)$ for the dominant Floquet harmonics generated inside the temporally modulated dielectric slabs of the optical device, under both forward and backward illumination. Panels show the field maps at the harmonic frequencies: 
(a) $\omega_{-1} = \omega_{0} - \Omega$, 
(b) $\omega_{0}$, 
(c) $\omega_{+1} = \omega_{0} + \Omega$, 
(d) $\omega_{-2} = \omega_{0} - 2\Omega$, and 
(e) $\omega_{+2} = \omega_{0} + 2\Omega$. 
For each harmonic, the left and right subpanels correspond to forward and backward incidence, 
respectively. All maps for the same harmonic index $m$ share a common linear colormap scale, 
enabling direct comparison of forward/backward amplitudes and illustrating the direction-dependent harmonic content produced by the time-modulated slabs.}
\label{fig:figS3}
\end{figure}

\begin{figure*}[!t]
\centering
\includegraphics[width=0.99\linewidth]{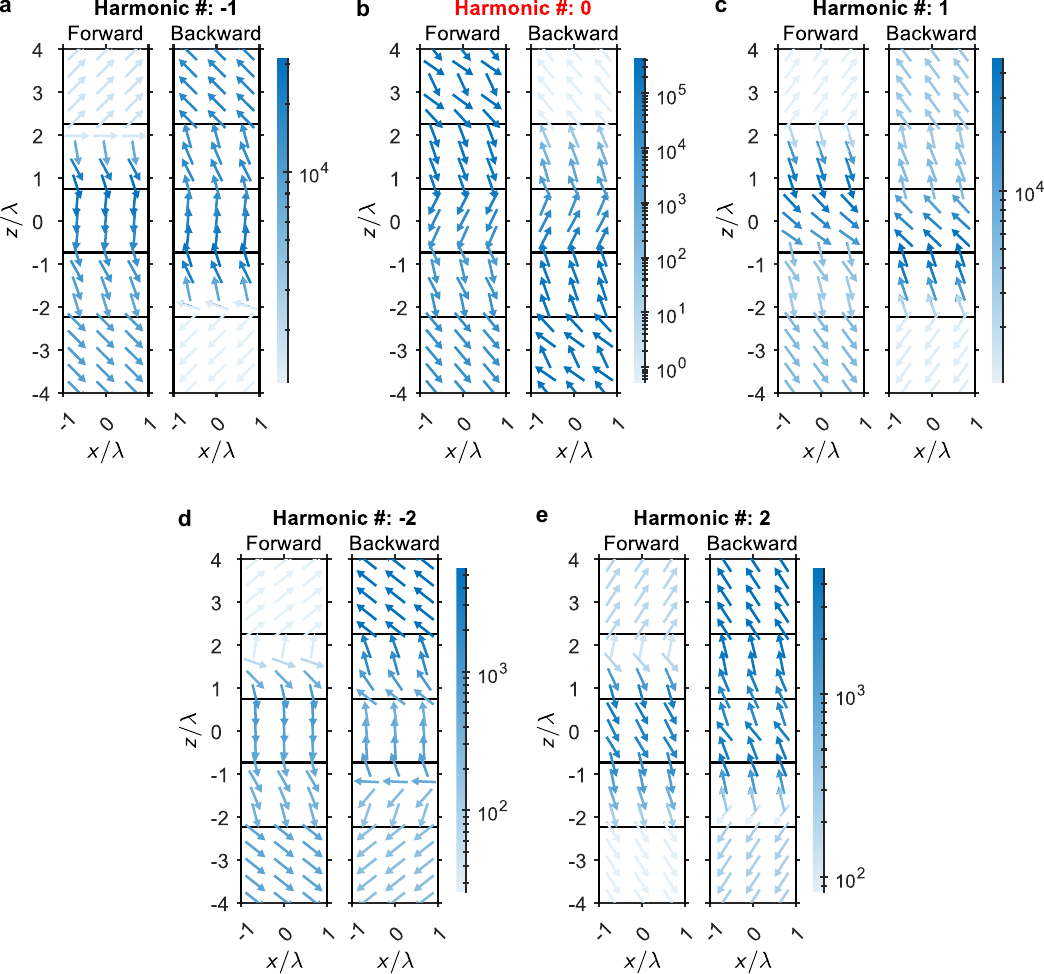}
\caption{Power-flow distributions of individual Floquet harmonics in the optical device. Time-averaged Poynting vector fields for the principal harmonics generated in the temporally modulated dielectric slabs of the optical device, shown under forward and backward illumination. Panels correspond to 
(a) $m=-1$, 
(b) $m=0$, 
(c) $m=+1$, 
(d) $m=-2$, and 
(e) $m=+2$, 
where each harmonic occurs at $\omega_{m} = \omega_{0} + m\Omega$. For every harmonic, the left and right subpanels display the forward- and backward-incidence power-flow patterns, respectively. All panels use a common logarithmic colormap scale to highlight the broad dynamic range of power amplitudes and to reveal subtle directional differences in harmonic energy transport.}
\label{fig:figS4}
\end{figure*}

Figures~\ref{fig:figS3} and \ref{fig:figS4} further quantify how electromagnetic power is distributed among Floquet harmonics within the multilayer device. The incident power per simulation period is normalized to \(P_{\rm inc}=1~\mathrm{W/m}\), and all computed values are referenced to this level. These results show that temporal modulation continuously feeds energy from the fundamental harmonic into multiple sidebands, while the hyperbolic slab, being dispersive, selectively enhances or suppresses specific harmonics depending on their longitudinal wavenumbers. As a result, the internal power flow strongly depends on the incidence direction, which is essential for the nonreciprocal transmission observed in the main text.

Supplementary Video~1 visualizes the time-domain evolution of the magnetic-field component 
$H_y(x,z,t)$ at the operating frequency $\omega_0$, corresponding to the zeroth-order 
Floquet harmonic, under both forward and backward illumination of the optical device. 
To enhance the perceptibility of directional differences in transmitted amplitude, 
the video employs a nonlinear $\mu$-law colormap with $\mu = 10$. This dynamic-range 
compression makes the contrast between forward and backward transmission more visible 
than a linear scale would allow. The video therefore complements Figs.~\ref{fig:figS3} and~\ref{fig:figS4} by 
providing a continuous-time depiction of how the main harmonic propagates through the 
modulated slabs and hyperbolic core, revealing the asymmetric field amplitudes 
associated with nonreciprocal transmission.

\begin{figure}[!t]
\centering
\includegraphics[width=0.99\linewidth]{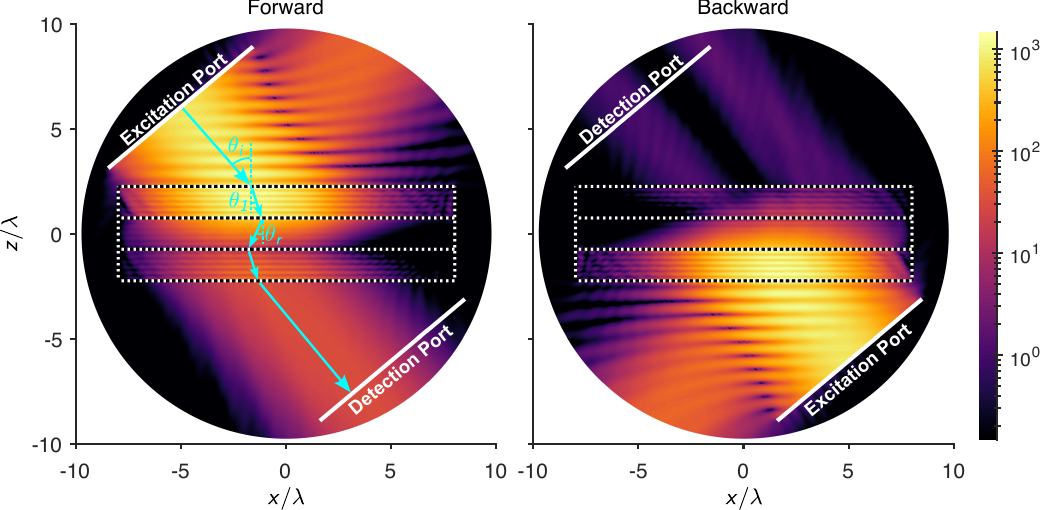}
\caption{ Gaussian-beam negative refraction through the optical device including the time-varying dielectric slabs. Magnetic-field intensity distributions $|\mathbf{H}|^{2}$ at the operating frequency $\omega_{0}$ for forward and backward Gaussian-beam illumination at an incident angle of 
$\theta_{i} = 40^{\circ}$. The dotted black--white lines denote the boundaries of the optical structure, in which the top and bottom regions are the temporally modulated dielectric slabs and the central region is the hyperbolic medium. Both surface plots use a common logarithmic color scale spanning a 40-dB dynamic range. Cyan arrows indicate the refraction trajectories, highlighting the negative-refraction behavior inside the hyperbolic core. From the simulated field patterns, the refraction angle at the air--modulated-slab interface is $\theta_{1} = 18.7^{\circ}$, and the refraction angle at the modulated-slab--hyperbolic interface is 
$\theta_{r} = -26.7^{\circ}$, confirming the backward-wave propagation intrinsic to the hyperbolic region.}
\label{fig:figS5}
\end{figure}

Figure~\ref{fig:figS5} highlights the spatial field profiles associated with forward and backward illumination, again assuming an incident power of \(P_{\rm inc}=1~\mathrm{W/m}\). The key observation is that the transmitted beam bends toward the ``negative'' side of the normal, confirming the occurrence of negative refraction in the presence of the temporal modulation. In addition, the forward and backward incidence panels show distinctly different refracted trajectories and field intensities. This directional asymmetry arises from the time modulation, which breaks time-reversal symmetry and leads to different modal interference pathways when the excitation direction is reversed.

\begin{figure}[!t]
\centering
\includegraphics[width=0.95\linewidth]{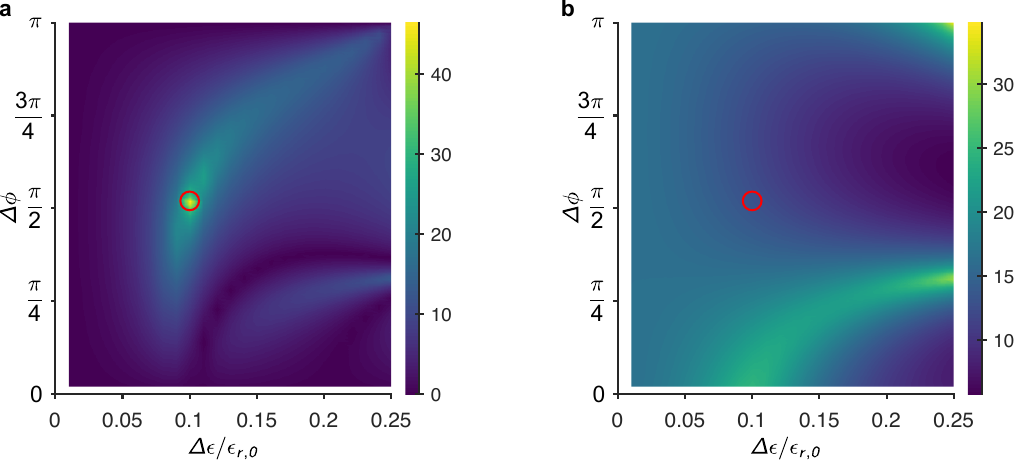}
\caption{
Influence of modulation parameters on the performance of the optical device. (a) Isolation of the optical device as a function of the modulation depth $\Delta\epsilon / \epsilon_{r,0}$ and the phase difference $\Delta\phi = \phi_{2} - \phi_{1}$ between the two temporally modulated dielectric slabs, shown on a dB scale. (b) Forward-mode insertion loss versus the same modulation parameters, also presented in dB. In both panels, the red circle indicates the selected operating point of the device and its corresponding performance, illustrating the trade-off between isolation and insertion loss across the modulation-parameter space.}
\label{fig:figS6}
\end{figure}

Finally, Fig.~\ref{fig:figS6} investigates the dependence of device performance on the modulation phase difference \(\phi_2-\phi_1\) applied to the two time-varying slabs. The insertion loss is defined as
\[
\mathrm{IL}=-10\log_{10}(T_f),
\]
with \(T_f\) denoting the forward transmittance. As shown in Fig.~\hyperref[fig:figS6]{\ref*{fig:figS6}(a)}, the isolation is maximized at
\[
\phi_2-\phi_1\approx \frac{\pi}{2}+\frac{\pi}{50}.
\]
although the insertion loss is not minimized. Thus, we can more clearly feel the trade-off between isolation and insertion loss. This optimal point lies slightly above the \(\pi/2\) quadrature condition predicted by ideal temporal coupled-mode theory~\cite{chegnizadeh_non-reciprocity_2020}. The shift arises from finite-thickness effects and the multimode interaction mediated through the dispersive hyperbolic core, which introduce additional internal phase accumulation. Consequently, the effective interference condition governing nonreciprocal transmission is achieved at a phase that is marginally offset from the idealized resonance limit.

\subsection{Power exchange calculations in the presence of time-varying media}

We begin with the definition of the instantaneous Poynting vector,
$\mathbf{S}=\mathbf{E}\times\mathbf{H}$, and apply the vector identity
$\nabla\!\cdot(\mathbf{E}\times\mathbf{H})
=\mathbf{H}\!\cdot(\nabla\times\mathbf{E})
-\mathbf{E}\!\cdot(\nabla\times\mathbf{H})$.
Using Maxwell’s curl equations,
$\nabla\times\mathbf{E}=-\partial_t\mathbf{B}$
and
$\nabla\times\mathbf{H}=\mathbf{J}_{\rm free}+\partial_t\mathbf{D}$,
we obtain
\begin{equation}
\nabla\!\cdot\mathbf{S}
+\mathbf{E}\!\cdot\partial_t\mathbf{D}
+\mathbf{H}\!\cdot\partial_t\mathbf{B}
=
-\,\mathbf{E}\!\cdot\mathbf{J}_{\rm free}.
\label{eq:poynting-general}
\end{equation}

Assuming an instantaneous scalar material law $\mathbf{D}=\epsilon(t)\mathbf{E}$, we may evaluate
\[
\mathbf{E}\!\cdot\partial_t\mathbf{D}
=
\mathbf{E}\!\cdot\big[\dot{\epsilon}(t)\mathbf{E}
+\epsilon(t)\partial_t\mathbf{E}\big].
\]
Meanwhile,
\[
\frac{\partial}{\partial t}\!\left(\tfrac12\,\mathbf{E}\!\cdot\mathbf{D}\right)
=
\tfrac12\,\mathbf{E}\!\cdot\dot{\epsilon}(t)\mathbf{E}
+
\mathbf{E}\!\cdot\epsilon(t)\partial_t\mathbf{E}.
\]
Subtracting the latter from the former yields the important identity
\[
\mathbf{E}\!\cdot\partial_t\mathbf{D}
=
\frac{\partial}{\partial t}\!\left(\tfrac12\,\mathbf{E}\!\cdot\mathbf{D}\right)
+
\tfrac12\,\mathbf{E}\!\cdot\dot{\epsilon}(t)\mathbf{E},
\]
which shows that $\mathbf{E}\!\cdot\partial_t\mathbf{D}$ decomposes into a stored-energy term and an
extra term that appears only when the permittivity is explicitly time dependent. Inserting this
result into \eqref{eq:poynting-general} gives the generalized Poynting theorem for a time-varying
dielectric:
\begin{equation}
\nabla\!\cdot\mathbf{S}
+
\frac{\partial}{\partial t}\!\left(
\tfrac12\,\mathbf{E}\!\cdot\mathbf{D}
+
\tfrac12\,\mathbf{H}\!\cdot\mathbf{B}
\right)
=
-\,\mathbf{E}\!\cdot\mathbf{J}_{\rm free}
-
\tfrac12\,\dot{\epsilon}(t)\,|\mathbf{E}(t)|^2.
\label{eq:poynting-timevarying}
\end{equation}
The final term on the right represents the per-unit-volume pump-field power exchange
\[
p_\epsilon(t)\;\triangleq\;\tfrac12\,\dot{\epsilon}(t)\,|\mathbf{E}(t)|^2,
\]
which acts as a source (if positive) or sink (if negative) depending on whether energy is being
injected into or extracted from the field by the modulation.

\subsubsection*{Fourier-series evaluation of the time-averaged power exchange}

To compute the time-averaged pump–field power transfer over one modulation period
$T=2\pi/\Omega$, we expand the field and material modulation in their Floquet–Fourier series. The
electric field is written as the real part of its multiharmonic expansion,
\[
\mathbf{E}(t)
=
\Re\!\left\{
\sum_{m\in\mathbb{Z}}\mathbf{E}_m\,e^{j(\omega_0+m\Omega)t}
\right\}
=
\tfrac12\sum_{m}
\big(
\mathbf{E}_m e^{j(\omega_0+m\Omega)t}
+\mathbf{E}_m^{*} e^{-j(\omega_0+m\Omega)t}
\big).
\]
Averaging $|\mathbf{E}(t)|^2$ over the fast optical oscillations removes all $2\omega_0$ terms and
produces a slowly varying envelope,
\[
|\mathbf{E}(t)|^2
\approx
\tfrac12\sum_{k}
\Re\!\big(C_k e^{jk\Omega t}\big),
\qquad
C_k=\sum_{n}\mathbf{E}_{n+k}\!\cdot\!\mathbf{E}_n^{*}.
\]
For a single-tone permittivity modulation,
\[
\epsilon(t)=\epsilon_{r,0}
+\Delta\epsilon\cos(\Omega t+\phi),
\]
we obtain
\[
\dot{\epsilon}(t)
=
\Re\!\left\{
A_{+1}e^{j\Omega t}
+
A_{-1}e^{-j\Omega t}
\right\},
\qquad
A_{+1}
=
\frac{j\Omega\Delta\epsilon}{2}e^{j\phi},
\quad
A_{-1}=A_{+1}^{*}.
\]

Multiplying the series for $\dot{\epsilon}(t)$ and $|\mathbf{E}(t)|^2$ and retaining only the
zero-frequency (DC) term gives the period-averaged pump power transfer,
\[
\langle p_\epsilon\rangle
=
\tfrac14\,\Re\!\big(A_{+1}C_{-1}+A_{-1}C_{+1}\big)
=
\frac{\Omega\Delta\epsilon}{4}
\Im\!\big\{e^{-j\phi}C_{+1}\big\}.
\]
Since $C_{+1}=\sum_{n}\mathbf{E}_{n+1}\!\cdot\!\mathbf{E}_n^{*}$, we arrive at the compact form
\[
\;
\langle p_\epsilon\rangle
=
\frac{\Omega\Delta\epsilon}{4}
\sum_{n\in\mathbb{Z}}
\Im\!\Big\{e^{-j\phi}
\,\mathbf{E}_{n+1}\!\cdot\!\mathbf{E}_n^{*}\Big\}.
\;
\]

Thus, for a single-tone temporal modulation, only adjacent harmonic pairs $(n,n+1)$ contribute to the
cycle-averaged pump–field energy exchange, and the net transfer depends on the relative phase 
$e^{-j\phi}$ and the quadrature relation between $\mathbf{E}_{n+1}$ and $\mathbf{E}_n$. A positive 
value of $\langle p_\epsilon\rangle$ corresponds to power delivered from the modulation to the field, 
while a negative value indicates net extraction of power from the field by the modulation.

\begin{figure}[!t]
\centering
\includegraphics[width=0.99\linewidth]{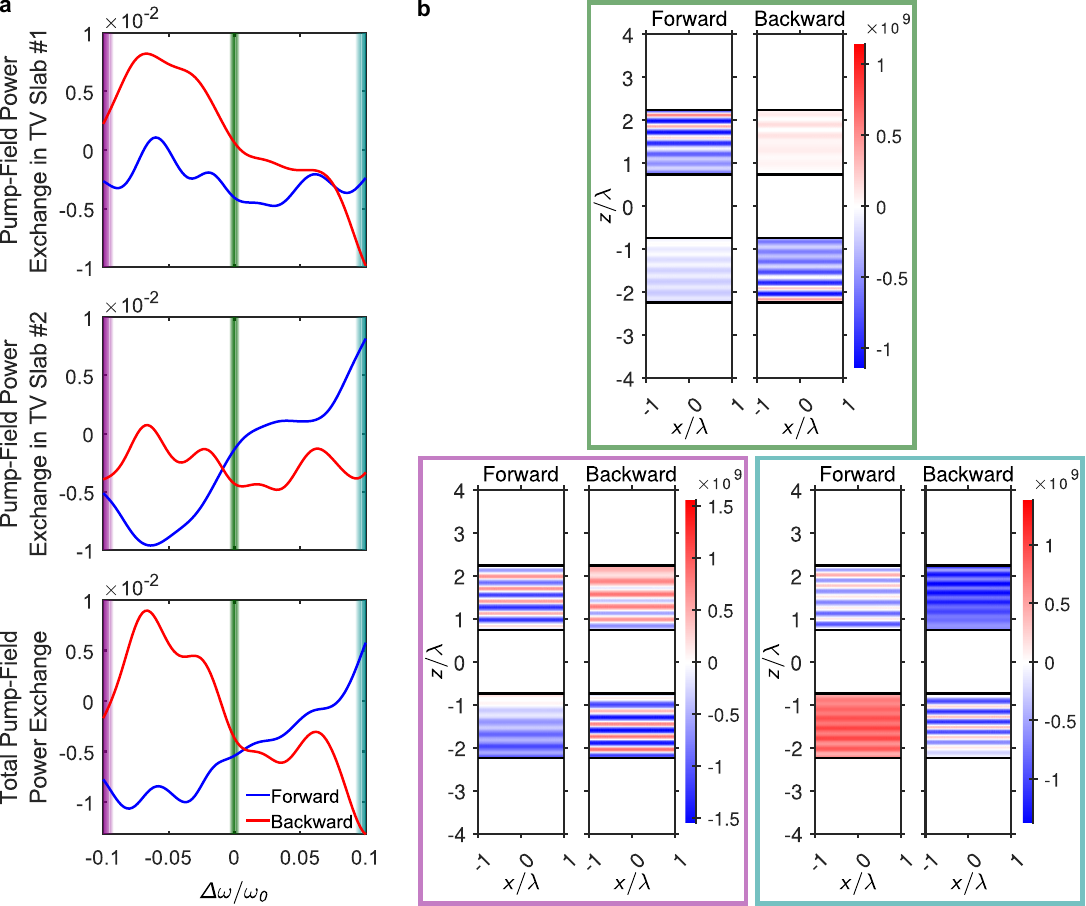}
\caption{Pump-field power exchange within the temporally modulated slabs of the optical device. (a) Pump-field power exchange in the two time-varying dielectric slabs (TV Slab~\#1 and TV Slab~\#2) as a function of the normalized detuning $\Delta\omega / \omega_{0}$, shown for both forward and backward illumination. The bottom panel displays the total pump-field power exchange obtained by summing the contributions from the two slabs. Shaded vertical regions highlight the three representative detuning values used in panel~(b). (b) Spatial distributions of the pump-field power exchange at detuning values of $-10\%$, $0\%$, and $+10\%$, again under forward and backward incidence. Each distribution is enclosed within a colored frame matching the shaded detuning regions in panel~(a). These profiles illustrate how the temporal modulation redistributes pump-field energy differently for the two incidence directions, and how this redistribution varies with detuning.}
\label{fig:figS7}
\end{figure}

Figure~\ref{fig:figS7} provides a direct illustration of these multiharmonic power-exchange mechanisms inside the
time-varying slabs. In Fig.~\hyperref[fig:figS7]{\ref*{fig:figS7}(a)}, the two upper panels show the individual contributions from the 
first and second modulated slabs (TV1 and TV2), plotted as the integrated pump–field interaction 
power within each slab. The bottom panel of Fig.~\hyperref[fig:figS7]{\ref*{fig:figS7}(a)} displays the resulting \emph{net} interaction 
power, obtained by summing the contributions of the two slabs. The results reveal that the overall 
pump–field interaction is different for forward and backward illumination, reflecting the 
direction-dependent harmonic content and modal interference entering each slab, even though the 
difference is not strongly asymmetric.

Figure~\hyperref[fig:figS7]{\ref*{fig:figS7}(b)} (right panels) complements this picture by presenting the spatial distribution of the 
local power-exchange density $p_\epsilon(x,z)$ across the structure. These maps show where, within 
the slabs, energy is transferred to or from the electromagnetic field for representative detunings 
$\Delta\omega/\omega_0=\{-10\%,\,0,\, +10\%\}$. The spatial patterns vary with both detuning and 
illumination direction, consistent with the multiharmonic coupling terms 
$\mathbf{E}_{n+1}\!\cdot\!\mathbf{E}_n^{*}$ that determine $\langle p_\epsilon\rangle$. Together, 
the panels of Fig.~\ref{fig:figS7} visualize how temporal modulation redistributes power among harmonics and across 
space, and how this redistribution differs for opposite propagation directions in the device.

\section{Microwave configuration}
\subsection{Matrix formulation for a hyperbolic medium covered by two time-modulated conductive sheets}

We consider a hyperbolic slab of thickness $L$ occupying $0<z<L$, with relative permittivity tensor
\[
\underline{\underline{\epsilon_r}}
=
\begin{pmatrix}
	\epsilon_t & 0          & 0 \\
	0          & \epsilon_t & 0 \\
	0          & 0          & \epsilon_z
\end{pmatrix}.
\]
Regions I ($z<0$) and III ($z>L$) are free space. A TM-polarized plane wave is incident from region~I
with angular frequency $\omega_0$ and incidence angle $\theta_{\rm inc}$.

The total $y$-polarized magnetic field in each region is expanded in Floquet harmonics at
$\omega_m=\omega_0+m\Omega$ as
\begin{align*}
\mathbf{H}^{\rm I}(x,z;t)
&=
\Bigl[
e^{-j k_{x,{\rm inc}}x} e^{-j k_{z,{\rm inc}}z} e^{j\omega_0 t}
+
e^{-j k_{x,{\rm inc}}x}
\sum_m r_m e^{+j k_{z,m}z} e^{j\omega_m t}
\Bigr]\hat{\mathbf{y}},\\[2pt]
\mathbf{H}^{\rm II}(x,z;t)
&=
e^{-j k_{x,{\rm inc}}x}
\sum_m\Bigl(
a_m e^{-j k_{z,m}^{\rm hyp}z}
+
b_m e^{+j k_{z,m}^{\rm hyp}z}
\Bigr)e^{j\omega_m t}\,\hat{\mathbf{y}},\\[2pt]
\mathbf{H}^{\rm III}(x,z;t)
&=
e^{-j k_{x,{\rm inc}}x}
\sum_m t_m e^{-j k_{z,m}z} e^{j\omega_m t}\,\hat{\mathbf{y}},
\end{align*}
where
\[
k_m=\frac{\omega_m}{c},\qquad
k_{x,{\rm inc}}=k_0\sin\theta_{\rm inc},\qquad
k_{z,{\rm inc}}=k_0\cos\theta_{\rm inc},
\]
\[
k_{z,m}=\sqrt{k_m^2-k_{x,{\rm inc}}^2},
\qquad
k_{z,m}^{\rm hyp}
=\sqrt{\epsilon_t\Bigl(k_m^2-\tfrac{k_{x,{\rm inc}}^2}{\epsilon_z}\Bigr)},
\]
and the angle associated with each harmonic in regions I and III is
\[
\theta_m=\arcsin\!\Bigl(\frac{k_{x,{\rm inc}}}{k_m}\Bigr),
\qquad |\theta_m|<\tfrac{\pi}{2}.
\]
For evanescent orders $k_{z,m}$ is purely imaginary with ${\rm Im}[k_{z,m}]<0$, and
$\theta_0=\theta_{\rm inc}$.

From the region-I expression it is clear that the incident field is
\[
\mathbf{H}_{\rm inc}(x,z;t)
=
e^{-j k_{x,{\rm inc}}x}e^{-j k_{z,{\rm inc}}z}e^{j\omega_0 t}\,\hat{\mathbf{y}}.
\]
Using the TM relation
\[
\mathbf{E}=\eta\,\mathbf{H}\times\frac{\mathbf{k}}{|\mathbf{k}|},
\]
we obtain the incident electric field
\[
\mathbf{E}_{\rm inc}(x,z;t)
=
\eta\,e^{-j k_{x,{\rm inc}}x}e^{-j k_{z,{\rm inc}}z}e^{j\omega_0 t}
\bigl(\cos\theta_{\rm inc}\,\hat{\mathbf{x}}-\sin\theta_{\rm inc}\,\hat{\mathbf{z}}\bigr).
\]

For each harmonic in free space (regions I and III), we write
\[
\mathbf{H}_m=H_m\hat{\mathbf{y}},\qquad
\mathbf{k}_m=k_{x,{\rm inc}}\hat{\mathbf{x}}+k_{z,m}\hat{\mathbf{z}}
=k_m(\sin\theta_m\hat{\mathbf{x}}+\cos\theta_m\hat{\mathbf{z}}),
\]
which leads to
\begin{align*}
k_{z,m}>0 &: \quad
\mathbf{E}_m
=
-\eta H_m\sin\theta_m\,\hat{\mathbf{z}}
+\eta H_m\cos\theta_m\,\hat{\mathbf{x}},\\[2pt]
k_{z,m}<0 &: \quad
\mathbf{E}_m
=
-\eta H_m\sin\theta_m\,\hat{\mathbf{z}}
-\eta H_m\cos\theta_m\,\hat{\mathbf{x}}.
\end{align*}
Inside the hyperbolic medium, using Maxwell’s equations with
$\underline{\underline{\epsilon_r}}=\mathrm{diag}(\epsilon_t,\epsilon_t,\epsilon_z)$ gives, for each
$\omega_m$,
\[
E_{x,m}^{\rm II}
=
\eta\,\frac{k_{z,m}^{\rm hyp}}{k_m\epsilon_t}H_m^{\rm II},
\qquad
E_{z,m}^{\rm II}
=
-\eta\,\frac{k_{x,{\rm inc}}}{k_m\epsilon_z}H_m^{\rm II},
\]
which directly leads to the series expressions in your original formulation.

\medskip
\noindent\textbf{Multiharmonic vector and operator notation.}
We now collect all harmonic amplitudes into column vectors (single underline) and introduce matrix
operators (double underline) in the harmonic space:
\[
\underline{r}
\triangleq (\ldots,r_{-2},r_{-1},r_0,r_{+1},r_{+2},\ldots)^{\mathsf T},
\quad
\underline{a},\,
\underline{b},\,
\underline{t}
\ \text{defined similarly},
\]
\[
\underline{1_0}
\triangleq (\ldots,0,0,1,0,0,\ldots)^{\mathsf T},
\]
\[
\underline{\underline{k}}
=\mathrm{diag}(\ldots,k_{-2},k_{-1},k_0,k_{+1},k_{+2},\ldots),\quad
\underline{\underline{k_z}}
=\mathrm{diag}(\ldots,k_{z,-2},k_{z,-1},k_{z,0},k_{z,+1},k_{z,+2},\ldots),
\]
\[
\underline{\underline{k^{\rm hyp}_z}}
=\mathrm{diag}(\ldots,k^{\rm hyp}_{z,-2},k^{\rm hyp}_{z,-1},k^{\rm hyp}_{z,0},
k^{\rm hyp}_{z,+1},k^{\rm hyp}_{z,+2},\ldots),
\]
and
\[
\underline{\underline{\theta}}
=\mathrm{diag}(\ldots,\theta_{-2},\theta_{-1},\theta_0,\theta_{+1},\theta_{+2},\ldots).
\]
Based on these, we define the diagonal “ratio” operator
\[
\underline{\underline{k^{\rm hyp}_z}}\;\underline{\underline{k}}^{-1}
=
\mathrm{diag}\!\Bigl(\ldots,\frac{k^{\rm hyp}_{z,-2}}{k_{-2}},\frac{k^{\rm hyp}_{z,-1}}{k_{-1}},
\frac{k^{\rm hyp}_{z,0}}{k_0},\ldots\Bigr),
\]
and the exponential and cosine operators
\[
e^{\pm j\underline{\underline{k_z}}z}
\triangleq
\mathrm{diag}(\ldots,e^{\pm j k_{z,-2}z},\ldots,e^{\pm j k_{z,+2}z},\ldots),
\]
\[
e^{\pm j\underline{\underline{k^{\rm hyp}_z}}z}
\triangleq
\mathrm{diag}(\ldots,e^{\pm j k^{\rm hyp}_{z,-2}z},\ldots,e^{\pm j k^{\rm hyp}_{z,+2}z},\ldots),
\]
\[
\cos(\underline{\underline{\theta}})
\triangleq
\mathrm{diag}(\ldots,\cos\theta_{-2},\cos\theta_{-1},\cos\theta_0,
\cos\theta_{+1},\cos\theta_{+2},\ldots).
\]

In this notation, the multiharmonic magnetic fields in the three regions can be written concisely as
\begin{align}
\underline{\mathbf{H}}^{\rm I}(x,z)
&=
e^{-j k_{x,{\rm inc}}x}
\Bigl[
e^{-j k_{z,{\rm inc}}z}\,\underline{1_0}
+
e^{+j\underline{\underline{k_z}}z}\,\underline{r}
\Bigr]\hat{\mathbf{y}},
\label{eq:H-mw-I}\\[2pt]
\underline{\mathbf{H}}^{\rm II}(x,z)
&=
e^{-j k_{x,{\rm inc}}x}
\Bigl[
e^{-j\underline{\underline{k^{\rm hyp}_z}}z}\,\underline{a}
+
e^{+j\underline{\underline{k^{\rm hyp}_z}}z}\,\underline{b}
\Bigr]\hat{\mathbf{y}},
\label{eq:H-mw-II}\\[2pt]
\underline{\mathbf{H}}^{\rm III}(x,z)
&=
e^{-j k_{x,{\rm inc}}x}
e^{-j\underline{\underline{k_z}}z}\,\underline{t}\,\hat{\mathbf{y}}.
\label{eq:H-mw-III}
\end{align}

Similarly, the tangential electric field components in each region follow from the relations above.
For the $x$-components we obtain
\begin{align*}
\underline{\mathbf{E}}_x^{\rm I}(x,z)
&=
\eta\,e^{-j k_{x,{\rm inc}}x}
\Bigl[
\cos\theta_{\rm inc}\,e^{-j k_{z,{\rm inc}}z}\,\underline{1_0}
-
\cos(\underline{\underline{\theta}})\,
e^{+j\underline{\underline{k_z}}z}\,\underline{r}
\Bigr]\hat{\mathbf{x}},\\[2pt]
\underline{\mathbf{E}}_x^{\rm II}(x,z)
&=
\eta\,e^{-j k_{x,{\rm inc}}x}\,
\frac{1}{\epsilon_t}\,
\underline{\underline{k^{\rm hyp}_z}}\;\underline{\underline{k}}^{-1}
\Bigl[
e^{-j\underline{\underline{k^{\rm hyp}_z}}z}\,\underline{a}
-
e^{+j\underline{\underline{k^{\rm hyp}_z}}z}\,\underline{b}
\Bigr]\hat{\mathbf{x}},\\[2pt]
\underline{\mathbf{E}}_x^{\rm III}(x,z)
&=
\eta\,e^{-j k_{x,{\rm inc}}x}
\cos(\underline{\underline{\theta}})\,
e^{-j\underline{\underline{k_z}}z}\,\underline{t}\,\hat{\mathbf{x}}.
\end{align*}
(The corresponding $z$-components follow analogously from the hyperbolic constitutive relations and
are not written explicitly here, since only the tangential components enter the sheet boundary
conditions.)

\medskip
\noindent\textbf{Time-modulated conductive sheets and boundary conditions.}
The two time-modulated conductive sheets at $z=0$ and $z=L$ are described by
\[
\mathbf{J}_s(t)=\sigma(t)\,\mathbf{E}_t(t),
\]
with $T$-periodic surface conductance
\[
\sigma(t)=\sum_n\sigma_n e^{jn\Omega t}.
\]
In harmonic space this corresponds to a Toeplitz matrix
\[
\underline{\underline{\sigma}}
=
\bigl[\sigma_{p-q}\bigr]_{p,q\in\mathbb{Z}},
\]
such that $(\underline{\underline{\sigma}}\,\underline{\mathbf{E}})_m
=\sum_n\sigma_{m-n}\mathbf{E}_n$. We denote by
$\underline{\underline{\sigma^0}}$ and $\underline{\underline{\sigma^L}}$ the Toeplitz matrices
associated with the sheets at $z=0$ and $z=L$, respectively.

The boundary conditions follow from continuity of the tangential electric field $E_x$ and from
\[
\hat{\mathbf{z}}\times\bigl(\mathbf{H}^{\rm up}-\mathbf{H}^{\rm down}\bigr)=\mathbf{J}_s(t)
=\sigma(t)\,\mathbf{E}_t(t),
\]
applied at $z=0$ and $z=L$ in the multiharmonic representation. Using the compact forms above, we
obtain the following matrix system for the unknown vectors
$\underline{r}$, $\underline{a}$, $\underline{b}$, and $\underline{t}$:
\begin{align*}
\cos(\underline{\underline{\theta}})\bigl(\underline{1_0}-\underline{r}\bigr)
&=
\frac{1}{\epsilon_t}\,
\underline{\underline{k^{\rm hyp}_z}}\;\underline{\underline{k}}^{-1}
\bigl(\underline{a}-\underline{b}\bigr),
\\[2pt]
\frac{1}{\epsilon_t}\,
\underline{\underline{k^{\rm hyp}_z}}\;\underline{\underline{k}}^{-1}
\Bigl[
e^{-j\underline{\underline{k^{\rm hyp}_z}}L}\,\underline{a}
-
e^{+j\underline{\underline{k^{\rm hyp}_z}}L}\,\underline{b}
\Bigr]
&=
\cos(\underline{\underline{\theta}})\,
e^{-j\underline{\underline{k_z}}L}\,\underline{t},
\\[4pt]
\underline{1_0}+\underline{r}-\underline{a}-\underline{b}
&=
\eta\,\underline{\underline{\sigma^0}}\,
\cos(\underline{\underline{\theta}})\bigl(\underline{1_0}-\underline{r}\bigr),
\\[2pt]
e^{-j\underline{\underline{k^{\rm hyp}_z}}L}\,\underline{a}
+
e^{+j\underline{\underline{k^{\rm hyp}_z}}L}\,\underline{b}
-
e^{-j\underline{\underline{k_z}}L}\,\underline{t}
&=
\eta\,\underline{\underline{\sigma^L}}\,
e^{-j\underline{\underline{k_z}}L}\,
\cos(\underline{\underline{\theta}})\,\underline{t}.
\end{align*}
By solving this linear system for $\underline{r}$, $\underline{a}$, $\underline{b}$, and
$\underline{t}$, we obtain all reflection and transmission harmonics of the microwave configuration.
In this notation, single underlines denote multiharmonic column vectors, and double underlines denote
matrix operators (including diagonal and Toeplitz matrices) acting in the harmonic space.

\subsection{Performance analysis of the microwave device in more detail}

\begin{figure}[!t]
\centering
\includegraphics[width=0.76\linewidth]{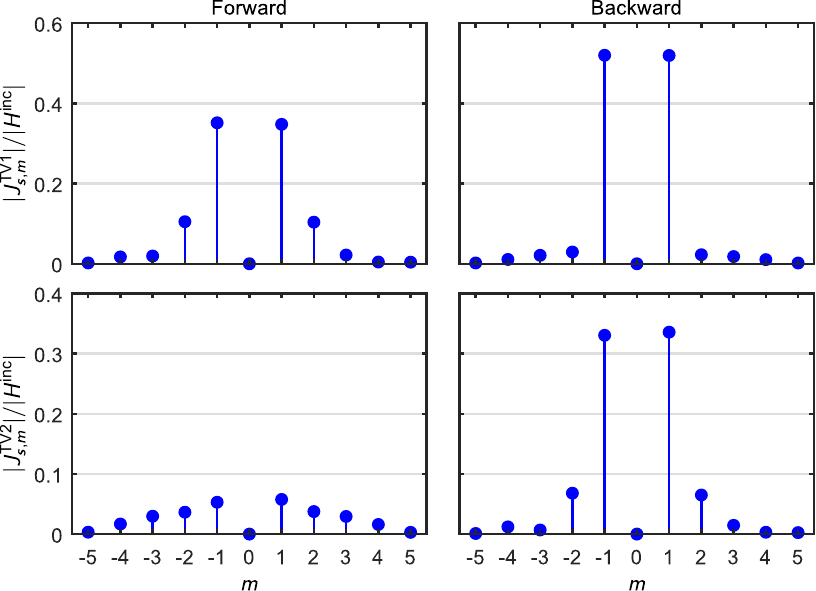}
\caption{Harmonic surface-current amplitudes generated on the time-modulated sheets of the microwave device. Stem plots of the magnitudes of the multiharmonic surface currents $|J^{\mathrm{TV1}}_{s,m}| / |H^{\mathrm{inc}}|$ and $|J^{\mathrm{TV2}}_{s,m}| / |H^{\mathrm{inc}}|$ induced on the two temporally modulated conductive sheets under forward and backward illumination. Each panel reports the amplitude of the surface current associated with harmonic index $m$, normalized to the incident magnetic-field amplitude. These distributions illustrate how temporal modulation of the sheet conductance generates direction-dependent sets of harmonic surface currents, providing the mechanism for nonreciprocal functionality in the microwave device.}
\label{fig:figS8}
\end{figure}

Figure~\ref{fig:figS8} illustrates the multiharmonic surface currents generated on the two time-modulated conductive sheets. Since each sheet is described through the Fourier expansion
\[
\sigma(t)=\sum_{n}\sigma_n e^{jn\Omega t},
\]
the induced surface current,
\[
\mathbf{J}_s(t)=\sigma(t)\mathbf{E}(t),
\]
naturally inherits a multiharmonic structure. Writing the fields in Floquet form,
\[
\mathbf{E}(t)=\sum_{m}\mathbf{E}_m e^{jm\Omega t},
\qquad
\mathbf{J}_s(t)=\sum_{m}\mathbf{J}_{s,m} e^{jm\Omega t},
\]
gives the harmonic-mixing relation
\[
\mathbf{J}_{s,m}=\sum_{n}\sigma_n\,\mathbf{E}_{m-n},
\]
which follows from the multiply--convolve property of Fourier transforms. Figure~\ref{fig:figS8} directly visualizes this mechanism: the two sheets generate distinct multiharmonic current patterns under forward versus backward excitation, and the redistribution among neighboring sidebands reflects the strength and phase of the temporal modulation.

\begin{figure}[!t]
\centering
\includegraphics[width=0.99\linewidth]{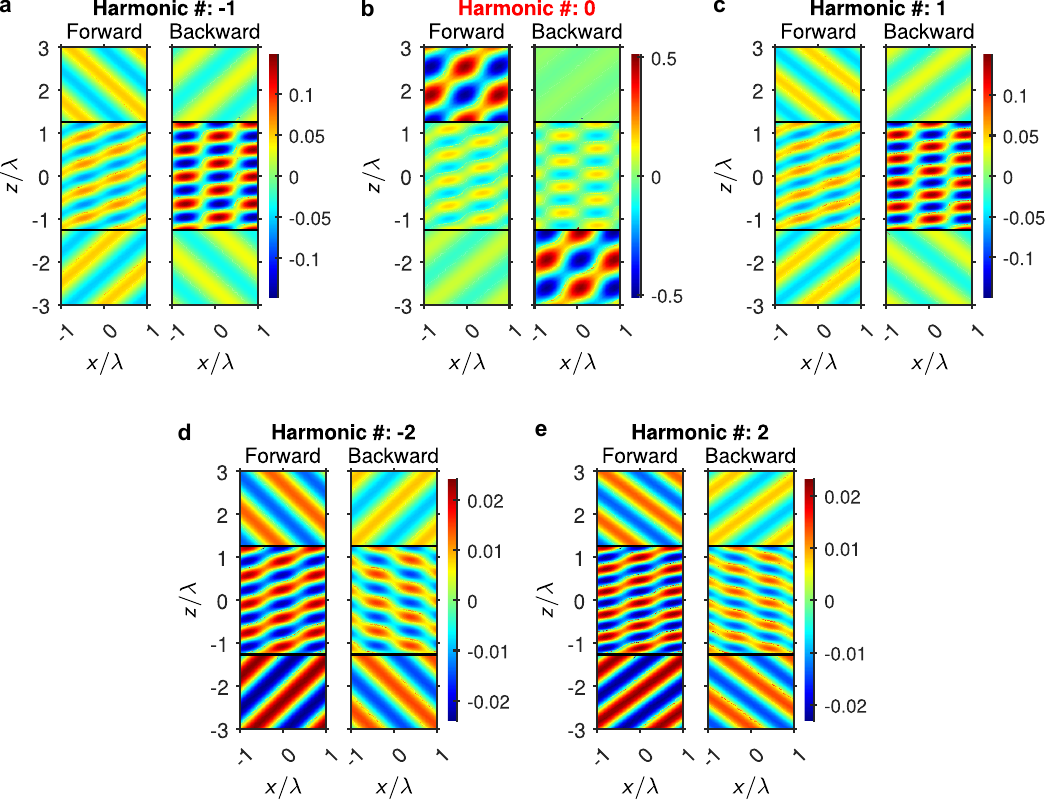}
\caption{Spatial magnetic-field profiles of Floquet harmonics generated by the time-modulated sheets in the microwave device. Spatial distributions of the magnetic field $H_{y}(x,z)$ associated with the Floquet harmonics produced when the two conductive sheets undergo sinusoidal modulation of their surface conductance. The harmonics displayed are 
(a) $m=-1$, 
(b) $m=0$, 
(c) $m=+1$, 
(d) $m=-2$, and 
(e) $m=+2$, 
corresponding to frequencies $\omega_{m} = \omega_{0} + m\Omega$. Each panel contains forward-illumination and backward-illumination results side by side, enabling a direct comparison of how the modulated sheets interact differently with waves incident from opposite directions. A consistent linear colormap is used for each harmonic order.}
\label{fig:figS9}
\end{figure}

\begin{figure}[!t]
\centering
\includegraphics[width=0.99\linewidth]{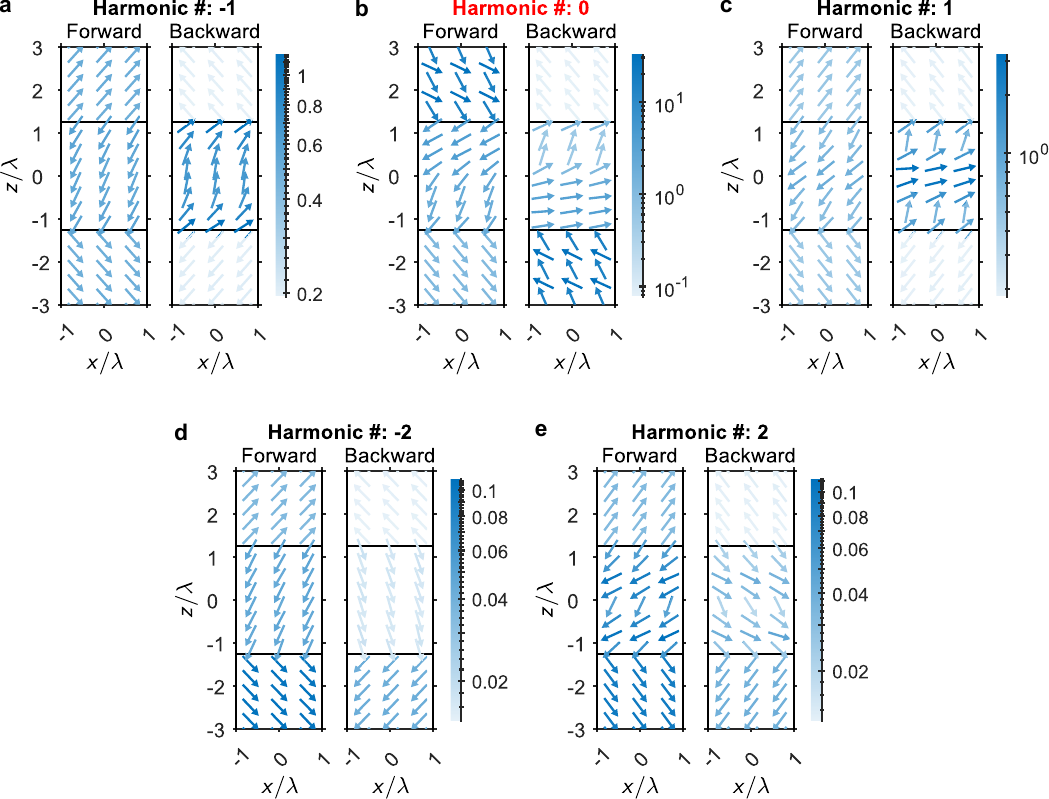}
\caption{Harmonic power-flow patterns produced by the time-modulated conductive sheets in the microwave device. Time-averaged Poynting vector distributions corresponding to the Floquet harmonics generated by the modulated sheets, shown for forward and backward incident waves. Panels depict the energy-flow fields for (a) \(m=-1\), (b) \(m=0\), (c) \(m=+1\), (d) \(m=-2\), and (e) \(m=+2\), with each harmonic occurring at \(\omega_{m} = \omega_{0} + m\Omega\). A common logarithmic color-scale is employed for each harmonic to clearly illustrate variations in power magnitude over a broad dynamic range. These asymmetric power-flow characteristics reveal the mechanism by which temporally varying sheet conductances induce nonreciprocal energy transport through the microwave structure.}
\label{fig:figS10}
\end{figure}

Figures~\ref{fig:figS9} and \ref{fig:figS10} show how the incident power (\(P_{\rm inc}=1~\mathrm{W/m}\)) is distributed among the Floquet harmonics inside the microwave structure. Similar to the optical device, the temporal modulation continuously injects energy into sidebands, while the hyperbolic slab selectively enhances or suppresses different harmonics depending on their longitudinal wavenumbers. Although the forward/backward contrast is somewhat milder than in the optical configuration, the internal harmonic power flow remains direction-dependent, further confirming the nonreciprocal behavior of the structure.

Supplementary Video~2 shows the time-domain evolution of the magnetic-field, 
$H_y(x,z,t)$, at the operating frequency $\omega_0$, again corresponding to the 
zeroth-order Floquet harmonic, for forward and backward excitation of the microwave 
device. As in the optical case, the video uses a nonlinear $\mu$-law colormap with 
$\mu = 10$ to compress the dynamic range and make the directional difference in 
transmitted amplitude visually prominent. Although the forward/backward contrast is 
milder than in the optical configuration, the nonlinear scaling allows the isolation 
to be easily perceived. This dynamic visualization complements Figs.~\ref{fig:figS9} and~\ref{fig:figS10} by 
showing how the fundamental harmonic evolves in time in response to the temporally 
modulated conductive sheets, thereby clarifying the origin of the device’s 
nonreciprocal behavior.

\begin{figure}[!t]
\centering
\includegraphics[width=0.99\linewidth]{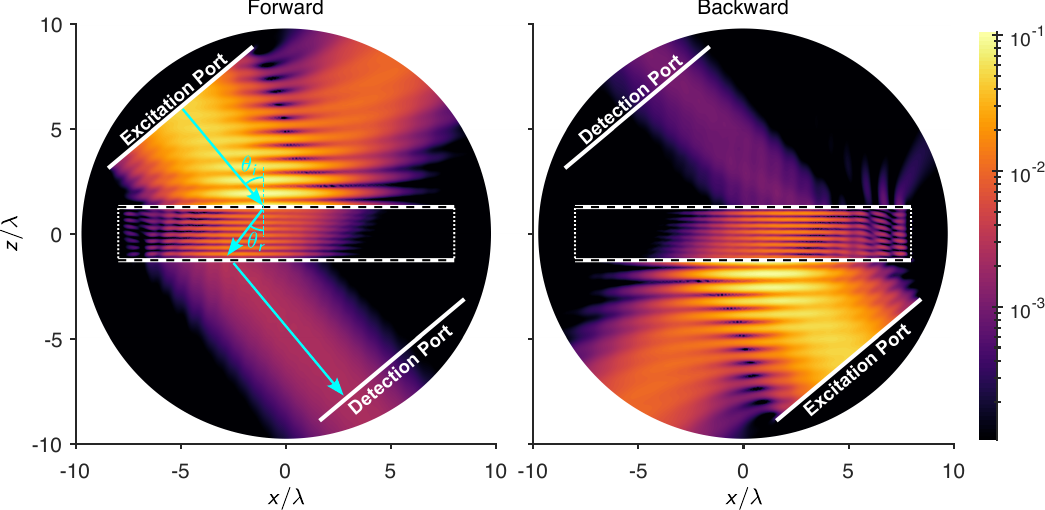}
\caption{Gaussian-beam negative refraction through the microwave device with time-modulated conductive sheets. Magnetic-field intensity distributions $|\mathbf{H}|^{2}$ at the operating frequency $\omega_{0}$ for Gaussian-beam illumination incident at $\theta_{i} = 40^{\circ}$ in the forward and backward directions. The thick dashed black-white lines mark the locations of the two temporally modulated conductive sheets enclosing the hyperbolic metamaterial slab (wire medium). Both field maps use a common logarithmic color scale spanning a 30-dB dynamic range. The refraction trajectory under forward incidence is denoted by cyan arrows, illustrating negative refraction within the hyperbolic layer. The refraction angle at the interface between air and the hyperbolic medium (coincident with the modulated sheets) is $\theta_{r} = -37.4^{\circ}$. The clear difference between forward and backward propagation reflects the asymmetric beam transmission induced by the temporal modulation of the conductive sheets.}
\label{fig:figS11}
\end{figure}

Figure~\ref{fig:figS11} presents the spatial field profiles for forward and backward illumination. The transmitted beam clearly refracts toward the negative side of the normal, demonstrating negative refraction in the microwave implementation. In addition, the field distribution inside the slab differs between the two excitation directions due to the distinct modal interference pathways generated by the temporally modulated sheets.

\begin{figure}[!t]
\centering
\includegraphics[width=0.95\linewidth]{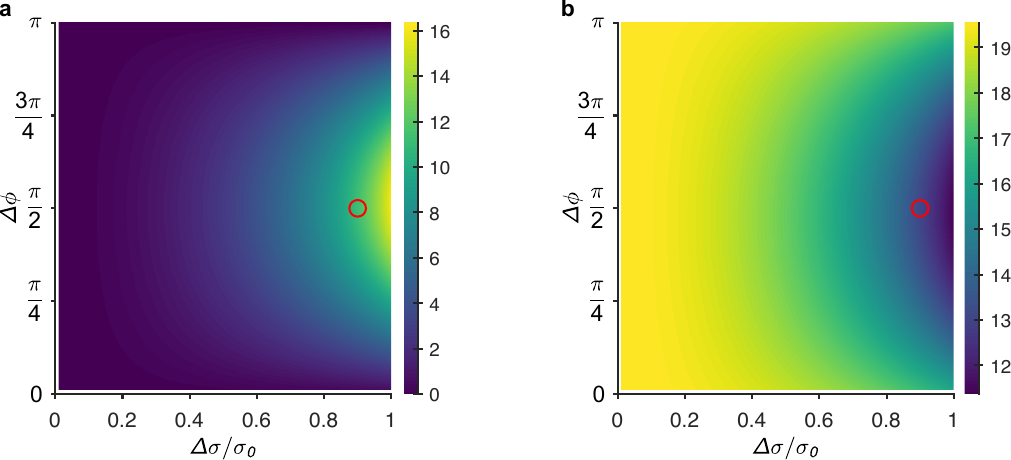}
\caption{Dependence of microwave-device performance on conductance modulation depth and phase offset. (a) Isolation of the microwave device as a function of the normalized conductance modulation depth $\Delta\sigma / \sigma_{0}$ and the phase difference $\Delta\phi = \phi_{2} - \phi_{1}$ between the two time-modulated conductive sheets. (b) Forward-mode insertion loss mapped over the same parameter space. In both panels, the red circle marks the chosen operating condition and the corresponding device performance. The color scales represent values in decibels.}
\label{fig:figS12}
\end{figure}

Finally, Figure~\ref{fig:figS12} examines how device performance varies with the modulation-phase difference between the two time-varying sheets. The insertion loss, previously defined in the corresponding optical section, exhibits its minimum near the quadrature point \(\phi_2-\phi_1 = \pi/2\), consistent with the theoretical condition predicted by temporal coupled-mode theory~\cite{chegnizadeh_non-reciprocity_2020}. Unlike the optical device, where finite-thickness and multimode coupling slightly shift the optimum phase, the microwave configuration reaches its best performance close to the ideal \(\pi/2\) value.

\subsection{Power exchange calculations in the presence of time-varying conductive sheets}

Since the microwave device contains no time-varying bulk dielectric regions, we begin directly from the standard Poynting theorem,
\begin{equation}
\nabla\!\cdot\mathbf{S}
+\frac{\partial}{\partial t}\!\left(\tfrac12\,\mathbf{E}\!\cdot\!\mathbf{D}
+\tfrac12\,\mathbf{H}\!\cdot\!\mathbf{B}\right)
= -\,\mathbf{E}\!\cdot\!\mathbf{J}_{\rm free},
\end{equation}
where the tangential sheet current density is included in $\mathbf{J}_{\rm free}$.  
For a time-modulated conductive sheet with instantaneous conductivity $\sigma(t)$, the instantaneous Ohmic power density is
\begin{equation}
p_\sigma(t)=\mathbf{E}(t)\!\cdot\!\mathbf{J}_{\rm free}(t)=\sigma(t)\,|\mathbf{E}(t)|^2.
\end{equation}
To determine how the modulation modifies power absorption, we compute the time average of $p_\sigma(t)$ over one modulation period $T=2\pi/\Omega$ using real-field (cosine) Fourier expansions with peak phasors and the convention $e^{+j\omega t}$.

\subsubsection*{Real-field Fourier representation and average dissipation}

The physical electric field is represented as the real part of a multiharmonic expansion,
\begin{equation}
\mathbf{E}(t)
=\Re\!\left\{\sum_{m\in\mathbb{Z}}\mathbf{E}_m e^{j(\omega_0+m\Omega)t}\right\}
=\tfrac12\sum_{m}\left(\mathbf{E}_m e^{j(\omega_0+m\Omega)t}
+\mathbf{E}_m^{*} e^{-j(\omega_0+m\Omega)t}\right),
\end{equation}
where $\mathbf{E}_m$ denote the peak complex amplitudes of the Floquet harmonics.  
The squared magnitude $|\mathbf{E}(t)|^2$ inherits an $\Omega$-periodic structure; after discarding the rapidly oscillating $2\omega_0$ components, it may be written as
\begin{equation}
|\mathbf{E}(t)|^2=\sum_{k\in\mathbb{Z}}\widetilde{C}_k\,e^{jk\Omega t},
\qquad
\widetilde{C}_k=\tfrac12\sum_{n}\mathbf{E}_{n+k}\!\cdot\!\mathbf{E}_n^{*},\qquad
\widetilde{C}_{-k}=\widetilde{C}_k^{*}.
\end{equation}

The sheet conductivity is expanded similarly. For single-tone modulation,
\begin{equation}
\sigma(t)=\sigma_0+\Delta\sigma\cos(\Omega t+\phi)
=
S_0 + S_{+1}e^{j\Omega t} + S_{-1}e^{-j\Omega t},
\qquad
S_0=\sigma_0,\quad S_{\pm 1}=\tfrac{\Delta\sigma}{2}e^{\pm j\phi}.
\end{equation}
Multiplying $\sigma(t)$ by $|\mathbf{E}(t)|^2$ yields another $\Omega$-periodic series.  
The time average $\langle p_\sigma\rangle$ corresponds to the DC coefficient, which is formed by all pairs of Fourier indices whose exponents sum to zero:
\begin{align}
\langle p_\sigma\rangle
&=
S_0\,\widetilde{C}_0
+ S_{+1}\widetilde{C}_{-1}
+ S_{-1}\widetilde{C}_{+1} \\
&=
\tfrac12\,\sigma_0\sum_{n}|\mathbf{E}_n|^2
+\tfrac12\,\Delta\sigma\,
\Re\!\Big\{e^{-j\phi}\sum_{n}\mathbf{E}_{n+1}\!\cdot\!\mathbf{E}_n^{*}\Big\}.
\end{align}
The first term is the familiar static Ohmic loss, while the second term describes the modulation-induced correction.  
Notably, this correction depends on the \emph{real} part of the cross-correlation between adjacent sidebands, in contrast with the time-varying-permittivity case (optical device), where the exchange term depended on the imaginary part of the corresponding correlation.

\subsubsection*{Surface-current formulation of the power exchange}

For a conductive sheet, the instantaneous surface power density is
\begin{equation}
p_s(t)=\mathbf{E}_t(t)\cdot\mathbf{J}_s(t),
\end{equation}
where $\mathbf{E}_t$ is the tangential electric field.  
Using real-field Floquet representations,
\begin{equation}
\mathbf{E}_t(t)=\Re\!\left\{\sum_{m}\mathbf{E}_m e^{j(\omega_0+m\Omega)t}\right\},
\qquad
\mathbf{J}_s(t)=\Re\!\left\{\sum_{m}\mathbf{J}_m e^{j(\omega_0+m\Omega)t}\right\},
\end{equation}
the product $\mathbf{E}_t\cdot\mathbf{J}_s$ contains many oscillatory terms, but only those with zero net modulation frequency contribute to the time average.  
This yields the compact relation
\begin{equation}
\langle p_s\rangle
=
\tfrac12\,\Re\!\left\{
\sum_{n\in\mathbb{Z}}
\mathbf{E}_n\cdot\mathbf{J}_n^{*}
\right\}.
\end{equation}
The factor $\tfrac12$ appears because the above uses peak phasors and real-part reconstructions.

\begin{figure}[!t]
\centering
\includegraphics[width=0.98\linewidth]{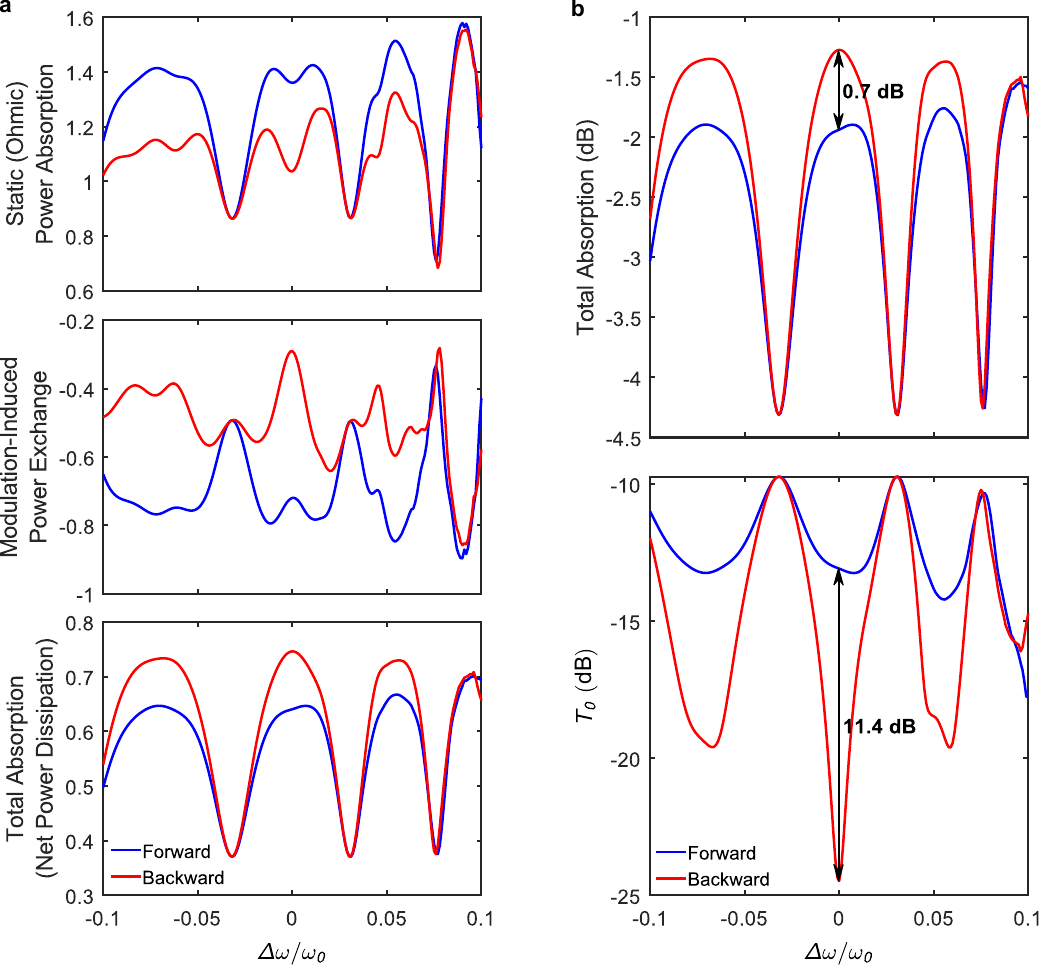}
\caption{Power dissipation and pump-driven power exchange in the time-modulated conductive sheets of the microwave device. (a) Contributions to power dissipation within the two temporally modulated conductive sheets as functions of the normalized frequency detuning $\Delta\omega / \omega_{0}$. The top panel shows the static (DC) ohmic absorption originating from the baseline sheet conductivity $\sigma_{0}$. The middle panel displays the modulation-induced power exchange associated with sideband coupling due to the time-varying conductance; this contribution may take positive or negative values. The bottom panel reports the total net absorbed power, combining both mechanisms. All power quantities are given in absolute values and normalized to the incident plane-wave power, and both forward and backward incidences are shown. (b) Total absorption (upper panel, in dB) and zeroth-order transmission $T_{0}$ (lower panel, in dB) as functions of normalized frequency detuning for forward and backward illumination.}
\label{fig:figS13}
\end{figure}

\medskip
Figure~\ref{fig:figS13} illustrates these mechanisms for the two time-modulated conductive sheets.  
In Fig.~\hyperref[fig:figS13]{\ref*{fig:figS13}(a)}, the decomposition into the static Ohmic term and the modulation-induced contribution clearly shows how harmonic mixing among adjacent Floquet components alters the net dissipation on each sheet.  
Finally, Fig.~\hyperref[fig:figS13]{\ref*{fig:figS13}(b)} compares forward and backward illumination. It can be observed that, for forward illumination, the conductive sheets exhibit lower absorption (0.7 dB less), leading to higher transmission, which results in an isolation of 11.4 dB.



\bibliographystyle{naturemag}  
\bibliography{RefsS}